\documentclass[aps,twocolumn,prd,superscriptaddress,showpacs,preprintnumbers,nofootinbib]{revtex4-1}

\usepackage{dsfont}
\usepackage{graphicx}
\usepackage{slashed}
\usepackage{amsmath,amssymb,amsthm,amsfonts}
\usepackage{color,rotating}
\usepackage{hyperref}
\usepackage[caption=false]{subfig}
 \def\Eq#1{Eq.~(\ref{#1})}
\def\Fig#1{Fig.~\ref{#1}} 

\def\0#1#2{\frac{#1}{#2}}
\def\eq#1{(\ref{#1})}
\def\dr{{D\!\llap{/}}\,}
\def\id{\mathds{1}}
\newcommand{\imag}{\text{i}}

\graphicspath{{./figures/}}

\begin{document}

\title{Chiral symmetry breaking in continuum QCD}

\author{Mario Mitter}
\affiliation{Institut f\"ur Theoretische
  Physik, Universit\"at Heidelberg, Philosophenweg 16, 69120
  Heidelberg, Germany} 

 \author{Jan M. Pawlowski} 
 \affiliation{Institut f\"ur Theoretische
  Physik, Universit\"at Heidelberg, Philosophenweg 16, 69120
  Heidelberg, Germany} 
\affiliation{ExtreMe Matter Institute EMMI, GSI, Planckstr. 1,
  D-64291 Darmstadt, Germany}

\author{Nils Strodthoff}
\affiliation{Institut f\"ur Theoretische
  Physik, Universit\"at Heidelberg, Philosophenweg 16, 69120
  Heidelberg, Germany}

\begin{abstract}
  We present a quantitative analysis of chiral symmetry breaking in
  two-flavour continuum QCD in the quenched limit. The theory is
  set-up at perturbative momenta, where asymptotic freedom leads to
  precise results. The evolution of QCD towards the hadronic phase is
  achieved by means of dynamical hadronisation in the non-perturbative
  functional renormalisation group approach.

  We use a vertex expansion scheme based on gauge-invariant operators
  and discuss its convergence properties and the remaining systematic
  errors. In particular we present results for the quark propagator,
  the full tensor structure and momentum dependence of the quark-gluon
  vertex, and the four-fermi scatterings.
\end{abstract}
\maketitle

\section{Introduction}

The understanding of the hadron spectrum as well as the phase
structure of QCD at finite temperature and density are very important
and long-standing problems.  Already a qualitative access to the
hadron spectrum beyond low-lying resonances and the phase structure at
large densities requires a quantitative hold on competing fluctuations
as well as the phenomena of dynamical chiral symmetry breaking and
confinement. 

Building on previous studies \cite{Braun:2008pi,Braun:2009gm}, this
work together with a related qualitative study of the unquenched
system in \cite{Braun:2014ata} provides the foundation for achieving this
goal.  The present work and \cite{Braun:2014ata} are first works within a
collaboration (fQCD) aiming at a quantitative functional
renormalisation group framework for QCD \cite{FRG-QCD}. While the
correct implementation of relative fluctuation scales is not required
to reproduce the thermodynamic properties of QCD at vanishing chemical
potential \cite{Herbst:2013ufa}, it will become increasingly important
at finite chemical potential. As was detailed in
\cite{Helmboldt:2014iya} at the example of quantum/thermal and density
fluctuations, mismatches in thermal/density fluctuation scales
inevitably lead to large systematic errors at finite chemical
potential. This is particularly important for the question of the
potential critical endpoint in the QCD phase diagram.

\begin{figure*}
  \centering \subfloat[Yang-Mills FRG gluon dressing function,
  \eq{eq:glueprop_dressing}, taken from \cite{Fischer:2008uz,FP} in
  comparison to quenched lattice data
  \cite{Bowman:2004jm,Sternbeck:2006cg}.\hfill\textcolor{white}{.}]{
\includegraphics[width=0.48\textwidth]{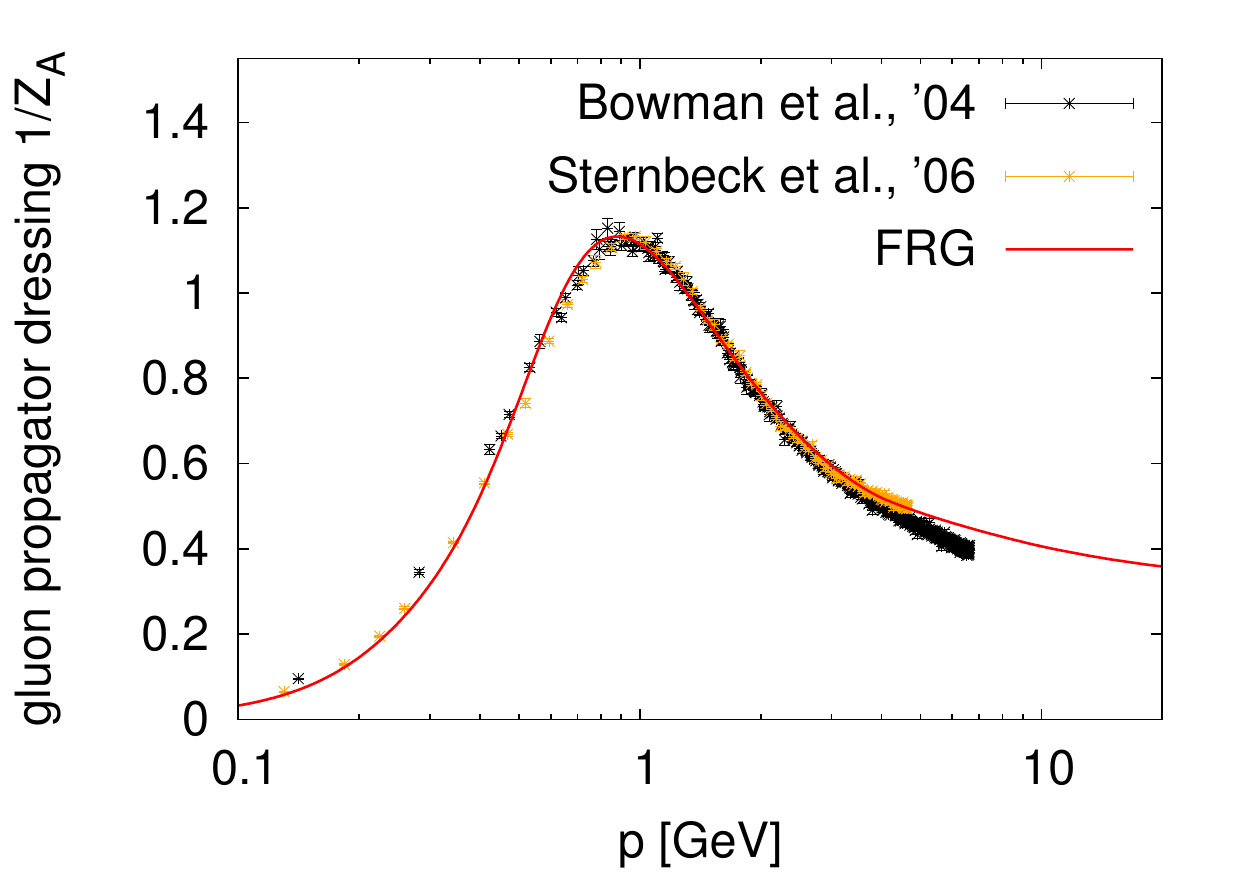}
    \label{fig:gluonprop}} \hfill \subfloat[Our
    result for quark propagator dressing
    functions, \eq{eq:quarkprop_dressing}, in comparison to quenched
    lattice results \cite{Bowman:2005vx} and mass function in
    GeV.\hfill\textcolor{white}{.}]{\includegraphics[width=0.48
\textwidth]{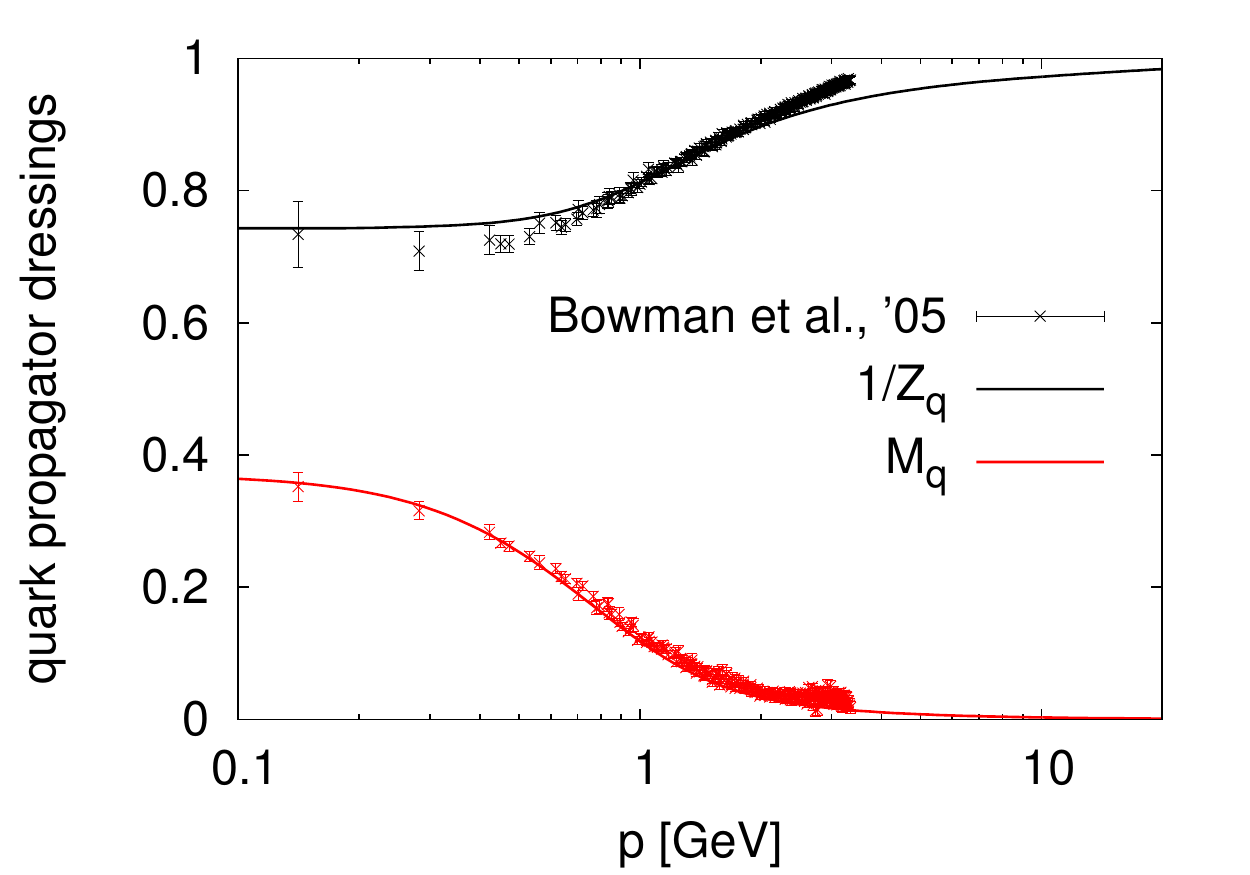}
\label{fig:quarkpropagator_msp}}
\caption{FRG results in comparison to lattice
    QCD. Dimensionful quantities in Bowman et al.,
    \cite{Bowman:2004jm,Bowman:2005vx}, are rescaled by a factor $0.91$
    in both plots to match the scales of Sternbeck et al.,
    \cite{Sternbeck:2006cg} and the FRG results,
    \cite{Fischer:2008uz,FP}.\hfill\textcolor{white}{.} }\label{fig:main_result}
\end{figure*}

Functional continuum approaches provide access to the mechanisms of
dynamical chiral symmetry breaking and confinement, as well as their
interrelation. Up to date, functional computations require larger or
smaller amounts of phenomenological input in the form of running
couplings, vertex models, or further low-energy parameters, see
\cite{Berges:2000ew,Pawlowski:2005xe,Gies:2006wv,Schaefer:2006sr,Braun:2011pp,
  Alkofer:2000wg,Roberts:2000aa,Fischer:2006ub,
  Fischer:2008uz,Binosi:2009qm,Maas:2011se,Boucaud:2011ug} and references
therein. In this work we present the first closed, self-consistent and
quantitative computation for quenched continuum QCD in the vacuum. A
prominent feature of this calculation is the lack of additional model
input, the computation only depends on the fundamental parameters of
QCD, the strong coupling $\alpha_s$ and the current quark masses which
are set at a large, perturbative momentum scales.  We implement a
systematic vertex expansion scheme that is fully capable of taking the
non-perturbative physics at low momenta into account. Gauge invariance
is implemented and tested in the form of modified Slavnov-Taylor
identities (mSTIs). In the present work we focus on the matter system
as one of the two subsectors of the full calculation. Using results
for the Yang-Mills propagators \cite{Fischer:2008uz,FP}, we solve the
matter sector in a quenched approximation and assess the quality of
our results in comparison to lattice QCD, see
Fig.~\ref{fig:main_result}. A separate analysis of the fully coupled
system is presented elsewhere.

The paper is organised as follows. In Sec.~\ref{sec:setup} we describe
our approach to QCD, in particular we briefly introduce the dynamical
hadronisation procedure within the functional renormalisation group
approach and describe the used truncation scheme. 
In Sec.~\ref{sec:results}, we present our results and comment on
the mechanism of chiral symmetry breaking in the light of our
investigations. Technical details on modified Slavnov-Taylor
identities and on our truncation can be found in the appendices.
\section{QCD with the Functional RG}
\label{sec:setup}
In quenched QCD there are no matter contributions to the gluon/ghost
correlation functions, since these contributions involve only diagrams
with closed quark loops.  Therefore, all
gluon/ghost correlation functions are given by those of the pure
Yang-Mills theory. 
Consequently, we use the functional renormalisation
group (FRG) results for Yang-Mills gluon and ghost propagators,
\cite{Fischer:2008uz,FP} in our calculation.

We perform a vertex expansion including a fully momentum-dependent
quark propagator and quark-gluon vertex as well as dynamically
generated four-fermi interactions in the matter sector and ghost-gluon
and three-gluon vertices in the glue sector. Furthermore we use gauge
invariance in the form of mSTIs to include higher quark-gluon
interactions as well as the four-gluon vertex. Mesonic interactions
are included via the mechanism of dynamical hadronisation, an RG-scale
dependent change of variables which constitutes an economic way to
take resonant structures in four-fermi interaction channels into
account. In the following two subsections we give a brief account of 
the FRG approach and our truncation scheme. A more
complete description of the truncation and a discussion of its stability is found in
Apps.~\ref{app:truncation} and \ref{app:results_stab}.

\subsection{Dynamical hadronisation in the functional renormalisation
  group}
\label{sec:frg}

The central object in the functional renormalisation group approach to
quantum field theory is the scale-dependent effective action
$\Gamma_k$. It generalises the effective action $\Gamma$, in the
spirit of the Wilsonian RG, by introducing a cutoff scale $k$ such
that $\Gamma_k$ includes only fluctuations from momentum modes with
momenta larger than $k$, see
\cite{Berges:2000ew,Pawlowski:2005xe,Gies:2006wv,Schaefer:2006sr,Braun:2011pp}
for QCD-related reviews. On a technical level this is achieved by
giving a momentum dependent 
mass to modes with momenta smaller than the scale $k$ by
means of an infrared regulator function $R_k$. In this way the
scale-dependent effective action $\Gamma_k$ interpolates between a
microscopic action, parameterised by a finite set of parameters, at
some large cutoff scale $k=\Lambda_\text{UV}$ and the full quantum
effective action in the limit $k\rightarrow 0$. The evolution of
$\Gamma_k$ with the RG-scale $k$ is described in terms 
an exact equation of one-loop structure, \cite{Wetterich:1992yh}
\begin{equation}
\label{eq:floweq}
\partial_t\Gamma_k=\frac{1}{2}\text{Tr}\,
\frac{1}{\Gamma_k^{(2)}+R_k}\partial_t R_k\,.
\end{equation}
Here $\Gamma_k^{(2)}$ denotes the second functional derivative with
respect to the fields, $t=\log(k/\Lambda)$ with some reference scale
$\Lambda$, and the trace includes a sum over all field species and
internal indices as well as a momentum-space integration. Note that
the flow equation \eq{eq:floweq} is one-loop exact, higher loop
corrections and non-perturbative effects are incorporated due the
presence of dressed, field-dependent propagators
$(\Gamma_k^{(2)}+R_k)^{-1}$. Flow equations for propagators or higher
order $n$-point functions are obtained by taking appropriate
functional derivatives of \eq{eq:floweq}. Despite its nature as an
exact equation, most practical applications require an Ansatz for the
scale-dependent effective action. Therefore, identifying the operators
that carry the relevant physical information is of utmost importance
for any quantitatively reliable solution of the flow
equation~(\ref{eq:floweq}).

\begin{figure}
  \includegraphics[width=0.48\textwidth]{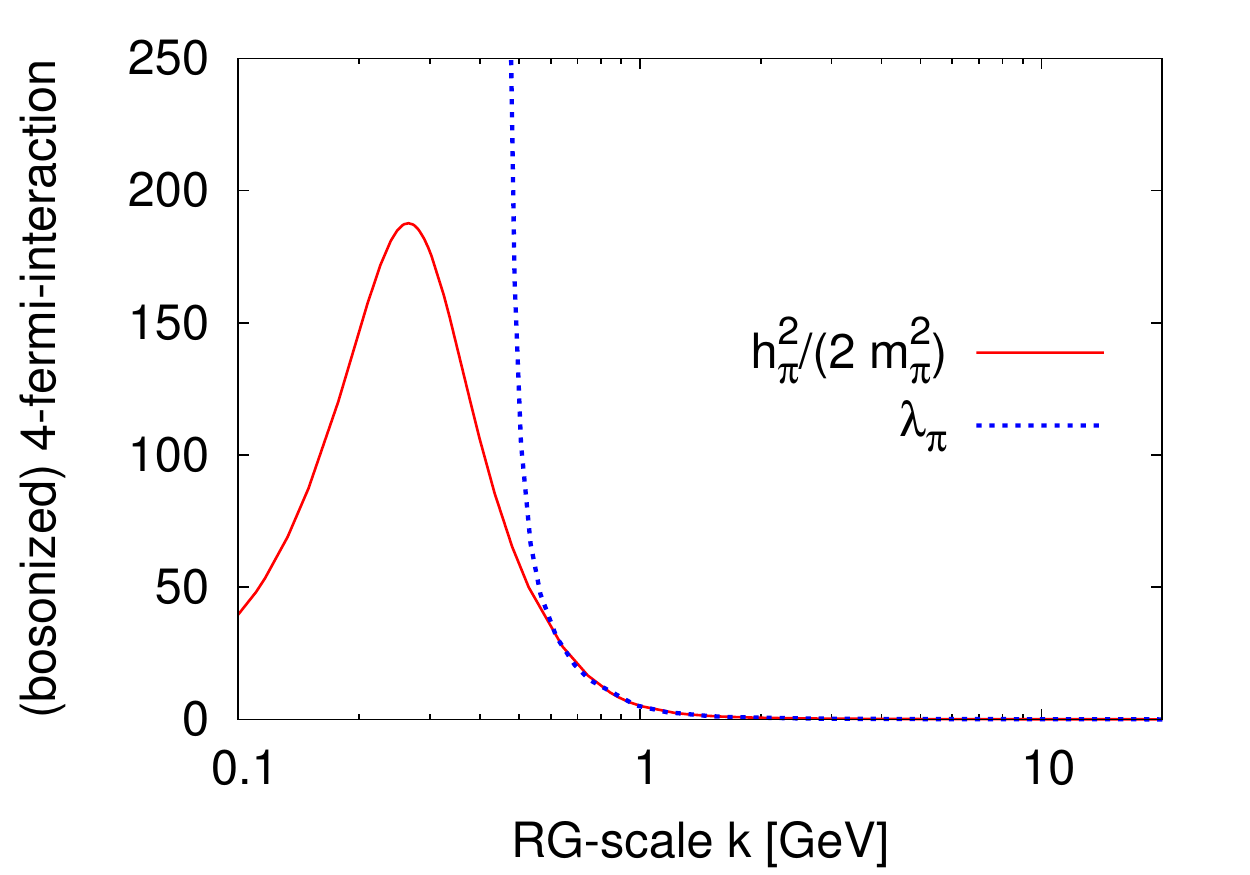}
  \caption{Dynamical hadronisation: four-fermi
      coupling ($\lambda_{\pi}$, see App.~\ref{trunc:4f})
      vs. corresponding coupling from dynamical hadronisation, $(k\,
      h_{\pi})^2/(2m_{\pi}^2)$, in a qualitative approximation.\hfill\textcolor{white}{.}}\label{fig:4fermirebos}
\end{figure}

Four-fermi interactions, e.g. in the scalar channel with coupling
$\lambda_{(\bar\psi\psi)^2}$, are created dynamically from two-gluon
exchange box diagrams that are proportional to $\alpha_s^2$. The
back-coupling of these four-fermi interactions on the system is
suppressed by additional powers of $\alpha_s^2$, for example its
contributions to the four-fermi system is of order $\alpha_s^4$.
However, as the strong running coupling, $\alpha_s$, becomes large
close to $\Lambda_{\text{QCD}}$, the suppression of the four-fermi
interactions is overcome and they start to grow large. 
As it becomes sufficiently large, the four-fermi
dynamics eventually become dominant and lead to a
four-fermi resonance. This resonance corresponds to the light pions as
pseudo--Nambu-Goldstone modes in the spontaneously broken phase. 
For even smaller momentum scales, quark
interactions exhibit dominant scatterings of these resonant momentum
channels. Hence, it is advantageous to describe these interactions in
terms of composite operators which is achieved via the
introduction of scale-dependent mesonic field operators
\cite{Gies:2001nw,Pawlowski:2005xe,Floerchinger:2009uf,Braun:2014ata}.
In the present work we follow the dynamical hadronisation procedure
set-up in \cite{Pawlowski:2005xe,Braun:2014ata}. 
In each renormalisation group step this leads e.g. to $\lambda_{\pi}\rightarrow h_\pi^2/(2
m_{\pi}^2)$ at vanishing $s$-channel momentum in the four-fermi channel. Here 
$\lambda_{\pi}$ corresponds to the exchange of pions with mass $m_\pi$
and Yukawa coupling $h_\pi$. This exact dynamical change
of field-variables avoids the numerically inconvenient singularity
shown in Fig.~\ref{fig:4fermirebos}, which is a consequence of
neglected momentum dependencies in the four-fermi interaction.
Additionally, it provides a smooth transition from QCD degrees of
freedom to constituent quarks and light mesons as low-energy effective
degrees of freedom. The resulting low-energy description in terms of a
quark-meson model introduces no model parameter dependence provided
the UV initial scale is chosen large enough
$\Lambda_\text{UV}\gg\Lambda_\text{QCD}$, see Sec.~\ref{sec:4fermi}
for an explicit demonstration.
Finally we want to stress that the restriction to such a small set
of low-energy degrees of freedom is justified by the comparably large
masses in the remainder of the spectrum of the strong interaction.
Since any of the hadrons can play a dynamical role only below about
$500$ MeV their fluctuations are strongly suppressed in any of the
loops by their comparably large mass, see also \cite{Braun:2014ata}.

\subsection{Truncation of effective action}
\label{sec:truncation}

\begin{figure*}
	\centering
	\includegraphics[width=0.98\textwidth]{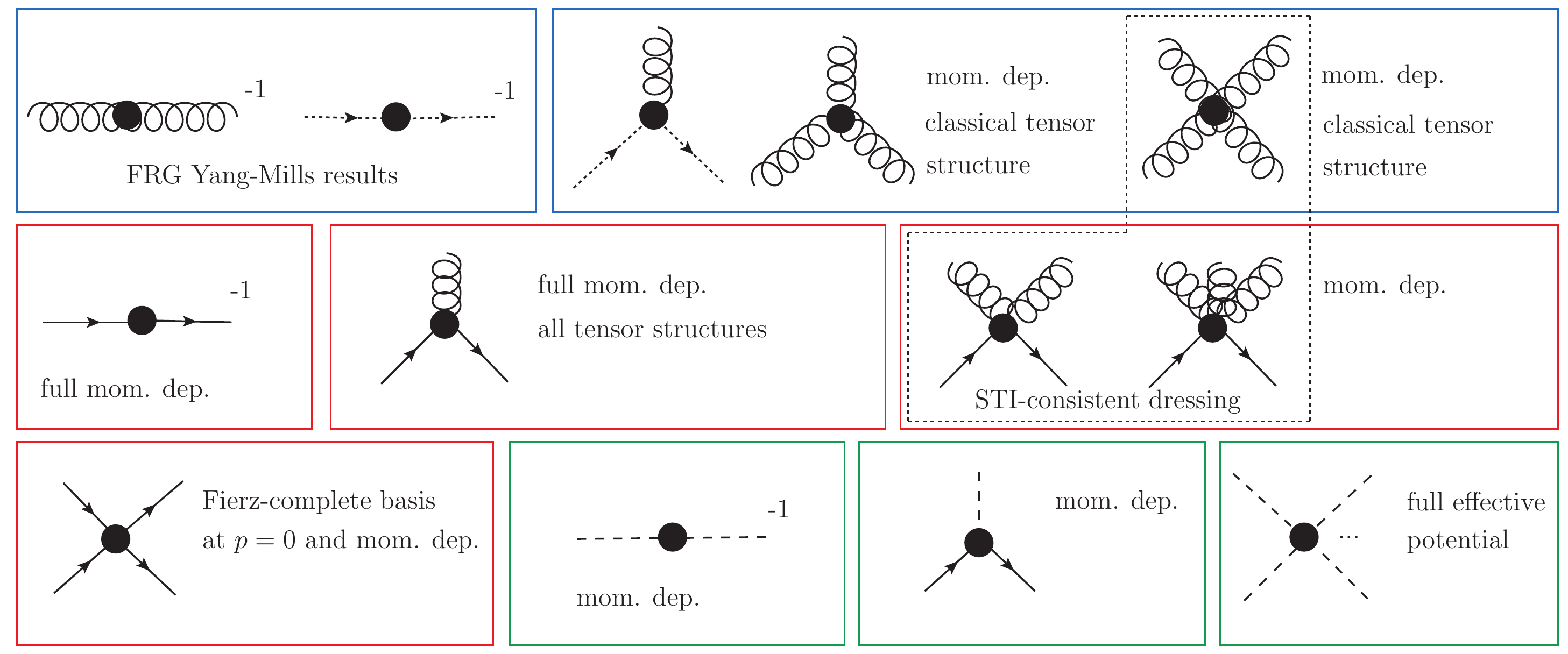}
	\caption{Pictorial description of our truncation for the effective action, see Fig.~\ref{fig:flow_I} for corresponding RG flows.}
	\label{fig:truncation}
\end{figure*}

In our truncation we consider the momentum dependence
  of all vertices which include at least one relevant or marginally
  relevant operator with the help of \cite{program}. 
  In the pure glue sector we calculate the ghost-
  and three-gluon vertices in single channel approximations including only
  the classical tensor structure. Moreover we use modified
  Slavnov-Taylor identities to fix the momentum dependence of
  the four-gluon vertex in this channel. This approximation is motivated
by results from other methods \cite{Pelaez:2013cpa,Blum:2014gna,Eichmann:2014xya,Binosi:2014kka,Gracey:2014mpa,Cyrol:2014kca},
which show non-trivial momentum-dependencies only in momentum regions
where the gluon sector already starts to decouple from the system.
The matter-glue coupling as the interface between the two subsectors
of the system is of crucial importance for the whole system. Therefore
we include the full momentum-dependence and all eight tensor
structures in the quark-gluon vertex.  Furthermore, there are two
exceptions from the RG relevance counting, in the sense that we
also include perturbatively irrelevant operators in our truncation.
Firstly, in the matter sector we include in addition the four-fermi
interactions, which are required for the description of chiral
symmetry breaking.  Secondly, for any non-classical operator which
shows a significant contribution in the flow, we identify the
corresponding gauge-invariant completion and include the resulting
higher order vertices in the flow equations.

The general vertex construction follows
\cite{Fischer:2009tn}. Suppressing the explicit RG-scale dependence we
parameterise
\begin{align}\nonumber 
\Gamma^{(n)}_{\Phi_1\cdots \Phi_n}(p_1,...,p_{n-1}) & \\[2ex] 
& \hspace{-1.8cm}= 
\bar \Gamma^{(n)}_{\Phi_1\cdots 
\Phi_n}(p_1,...,p_{n-1}) \prod_{i=1}^{n}\sqrt{\bar Z_{\Phi_i}(p_i)}\,, 
\label{eq:genvert}\end{align}
where we introduced a superfield 
\begin{align}\label{eq:Phi}
\Phi=(A_\mu\,,\, c\,,\,\bar c\,,\,q\,,\,\bar q, \vec
\pi\,,\,\sigma\,,...)
\end{align}
which subsums all dynamical degrees of freedom including the effective
low-energy fields generated by dynamical hadronisation. The tensor
kernel $\bar\Gamma^{(n)}_{\Phi_1\cdots\Phi_n}$ is expanded in a basis
of tensor structures ${\cal T}_{\Phi_1\cdots \Phi_n}^{(i)}$
\begin{align}\label{eq:tensorkernel} 
&\bar\Gamma^{(n)}_{\Phi_1\cdots 
\Phi_n}(p_1,...,p_{n-1})= \nonumber\\[1ex]
&\qquad \sum\limits_i z_{\Phi_1\cdots \Phi_n}^{(i)}(p_1,....,p_{n-1}) 
 {\cal T}_{\Phi_1\cdots \Phi_n}^{(i)}(p_1,...,p_{n-1}) \,.
\end{align}
Since the dressing functions $z_{\Phi_1\cdots
  \Phi_n}^{(i)}(p_1,....,p_{n-1})$ depend on our choice of $\bar
Z_{\Phi_i}$, the latter are at our disposal to give special properties
like RG-invariance in the perturbative regime to the former. If not
specified otherwise, we choose $\bar Z_{\Phi_i}(p)\equiv
Z_{\Phi_i,k}(p)$.  An important example are the classical vertices
with tensor structures ${\cal T}_{\rm class}$ present in the classical
action. We use
\begin{align}\label{eq:classT}
 {\cal T}_{{\rm class},\Phi_1\cdots \Phi_n}(p_1,\ldots,p_{n-1})=&
S^{(n)}_{\Phi_1\ldots
    \Phi_n}\Bigr|_{g=1}\,,
\end{align}
where $S^{(n)}$ denotes the appropriate $n$-th functional derivative
of the action. By setting $g=1$ in \eq{eq:classT}, the running coupling
is taken into account via the dressing functions 
$z_{\Phi_1\cdots \Phi_n}^{(i)}$ in \eq{eq:tensorkernel}. 
As a consequence of our choice for $\bar Z_{\Phi_i}(p)$, 
the $z_{\Phi_1\ldots\Phi_n,k\equiv 0}^{(1)}(p_1,\dots,p_{n-1})$ run like 
appropriate powers of the strong running coupling with the momenta $p_i$
in the perturbative regime.
The same holds for the RG-scale dependence of 
the $z_{\Phi_1\ldots \Phi_n}^{(1)}(0,\dots,0)$ at perturbative scales. 

In the following we discuss in some detail the constituents of our
Ansatz for the bosonised effective action of Landau-gauge QCD, which
are also summarised pictorially in Fig.~\ref{fig:truncation}. For a
more detailed description the reader is referred to
App.~\ref{app:truncation}. The stability of this truncation and 
the systematic errors are discussed in App.~\ref{app:results_stab}.\\[-2ex]

\emph{Glue sector}: As a consistent scale-setting both in the perturbative
and in the non-perturbative regime is crucial for our calculation, we use
YM FRG data \cite{Fischer:2008uz,FP} for both the gluon propagator and 
the ghost propagator, see Fig.~\ref{fig:gluonprop}. 
Here, we have matched our scale
to the corresponding lattice scales in \cite{Sternbeck:2006cg} via the peak position in the gluon
dressing function $1/Z_A$, which translates to
\begin{eqnarray}
 \alpha_s(20 \text{ GeV}) = 0.21\ .
\end{eqnarray}
The dressing functions of the Yang-Mills three-point functions,
$z_{\bar c A c},\, z_{A^3}$, are calculated momentum-dependently for a
single momentum channel. The four-gluon vertex is approximated using
the three-gluon vertex, see App.~\ref{trunc:3g}. This is a very good
approximation down to semi-perturbative momenta
\cite{Pelaez:2013cpa,Blum:2014gna,Eichmann:2014xya,Binosi:2014kka,Gracey:2014mpa,Cyrol:2014kca},
wheres deviations occur mostly for momenta where the glue gap implies
already decoupling.
\\[-2ex]

\emph{Matter sector}: We take into account the full momentum
dependence of the quark propagator, parameterised by its wavefunction
renormalisation $Z_q(p)$ and mass function $M_q(p)$, 
where we have for the current quark mass
\begin{eqnarray}
M_q(20\text{ GeV}) = 1.3 \text{ MeV}, 
\end{eqnarray}
see App.~\ref{trunc:qpr} for details. 
The treatment of the quark-gluon vertex is of
crucial importance for the whole system. Therefore we include the full
momentum dependence of all eight linearly independent tensor
structures ${\cal T}^{(i)}_{\bar q A q}$ of Landau gauge as described
in App.~\ref{trunc:qgl}. Additionally we perform a gauge-invariant
completion of any quark-gluon vertex 
tensor structure that contributes quantitatively, leading to the inclusion of
two-quark--two-gluon and two-quark-three-gluon vertices that are
approximated gauge-invariantly, see App.~\ref{trunc:qgl}.

In the four-fermi sector we take into account a Fierz-complete basis
of all ten tensor structures consistent with a $U(1)_V\times SU(2)_V$
symmetry, see App.~\ref{trunc:4f}, and approximate their momentum
dependence using a single ($s$-channel) momentum variable. As
discussed previously, we utilize dynamical hadronisation to effectively
remove the resonant $\sigma-\pi$ channel from the four-fermi tensor
structures via the inclusion of effective (quark-)meson
interactions. 

In the mesonic sector we include a scale-dependent mesonic
wavefunction renormalisation factor, $Z_\pi$, and a Yukawa interaction
between quarks and mesons, $h_\pi$. Additionally a scale-dependent
effective potential, $U(\rho)$ captures higher mesonic interactions in
the non-perturbative regime of spontaneously broken chiral symmetry,
see e.g. \cite{Schaefer:2004en}. This approximation has been shown to
be in quantitative agreement with the full momentum dependence
\cite{Helmboldt:2014iya}.

\begin{figure*}
  \subfloat[Coupling strength, \eq{eq:alphasAqq} and
  \eq{eq:quarkgluon-basis}, of quark-gluon vertex tensor components at
  symmetric point. Black: classical tensor structure, grey: chirally
  symmetric non-classical tensor structures, red: tensor structures
  that break chiral
  symmetry.\hfill\textcolor{white}{.}]{\includegraphics[width=0.48\textwidth]{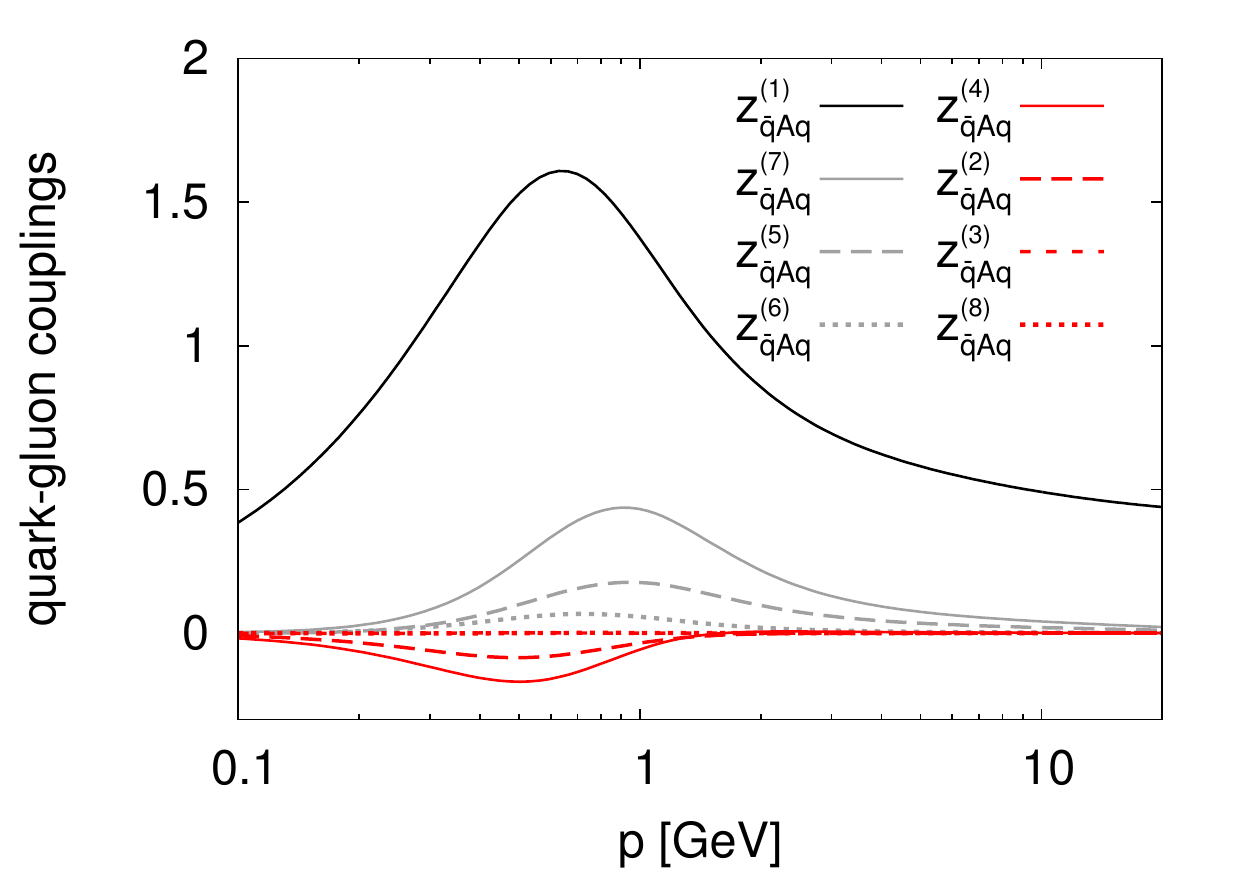}
                  \label{fig:quarkgluonvertex_components_msp}} \hfill
                \subfloat[Running couplings $\alpha_s$ at
                symmetric point $p_1^2=p_2^2=(p_1+p_2)^2$ from
                different vertices which takes the gluon gap into
                account analogous to \eq{eq:alphasAqq} for the
                quark-gluon vertex.  Critical gauge coupling for
                chiral symmetry breaking $\alpha_{\rm
                  crit}$.\hfill\textcolor{white}{.}]{\includegraphics[width=0.48\textwidth]{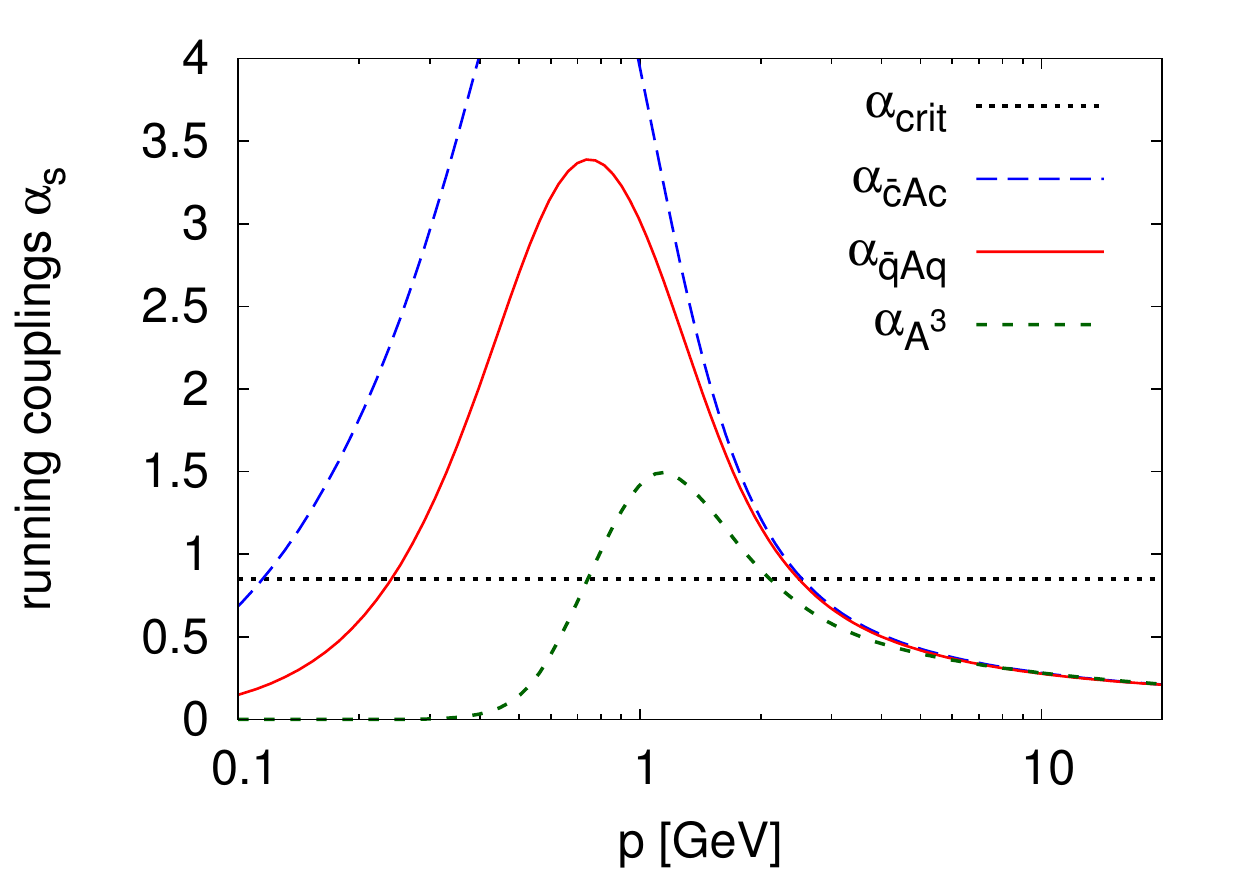}
\label{fig:alpha_msp}}
 \caption{ Strength of tensor structures and vertices.}\label{fig:running_couplings}
\end{figure*}

\section{Results}
\label{sec:results}

\subsection{Quark-gluon interactions}
\label{sec:quark}
We start by considering the quark-gluon vertex and focus in particular on 
additional non-classical tensor structures, as they are shown for the symmetric momentum configuration
$p_1^2=p_2^2 = p_3^2$ in 
Fig.~\ref{fig:quarkgluonvertex_components_msp}. To investigate their importance we calculate
the full momentum dependence of a basis for the transversally projected Landau-gauge quark-gluon vertex. 
The resulting eight tensor structures include the classical tensor structure, 
\begin{eqnarray}
 [{\cal T}^{(1)}_{\bar q A q}]^\mu=\gamma^\mu\,,
\end{eqnarray}
three further chirally symmetric tensors 
\begin{eqnarray}
 {\cal T}^{(5)}_{\bar q A q}\,,
  \qquad {\cal T}^{(6)}_{\bar q A q}\,,
  \qquad{\cal T}^{(7)}_{\bar q A q}\,,
\end{eqnarray}
and four tensors which break chiral symmetry 
\begin{eqnarray}
  {\cal T}^{(2)}_{\bar q A q}\,,
  \qquad{\cal T}^{(3)}_{\bar q A q}\,,
  \qquad{\cal T}^{(4)}_{\bar q A q}\,,
    \qquad{\cal T}^{(8)}_{\bar q A q}\,,  
\end{eqnarray}
listed explicitly in App.~\ref{trunc:qgl}. Each of the eight
tensor structures leads to a contribution in the effective action
that, if separated from the remainder of the action, 
violates gauge invariance. For example, the classical
tensor structure, $\gamma^\mu$, corresponds to the term $\bar q
\slashed A q$ in the effective action which is by itself not gauge
invariant. However, it appears as part of the gauge-covariant
derivative $\bar q \slashed D q$ which respects gauge
invariance. On the other hand, for the additional tensor structures, ${\cal
  T}^{(i)}_{\bar q A q}$, $i>1$, such a gauge-covariant completion is
not automatically included. A na\"ive inclusion of these tensor
structures alone would therefore violate gauge invariance in the form
of (modified) Slavnov-Taylor identities, see the discussion in
App.~\ref{app:mSTI}.

In the semi-perturbative, chirally symmetric regime we find the
gauge-invariant operator
\begin{eqnarray}\label{eq:STI_operator_I}
 \imag \sqrt{ 4 \pi \alpha_{s} }  \, \bar q\, \gamma_5
  \gamma_\mu \,\epsilon_{\mu\nu\rho\sigma} \{ F_{\nu\rho}\,,\,
  D_\sigma\} \,q\,,  
\end{eqnarray}
whose contribution to the term of $\mathcal{O}(\bar q A q)$ is
proportional to the tensor
\begin{eqnarray}
 \tfrac{1}{2}{\cal T}^{(5)}_{\bar q A q}+{\cal T}^{(7)}_{\bar q A q}\ .
\end{eqnarray}
Together with our results for the dressing functions, $z_{\bar q A
  q}^{(i)}(p,q)$ evaluated at $p^2=q^2=(p+q)^2$,
Fig.~\ref{fig:quarkgluonvertex_components_msp}, we conclude that this
is indeed the gauge-invariant operator that determines most of the
strength of the chirally symmetric non-classical tensor structures,
see also Fig.~\ref{fig:STI}.  Since the operator
in \eq{eq:STI_operator_I} contributes also to tensor structures in
higher vertices, namely the two-quark-two-gluon and two-quark-three-gluon
interactions, we include these as well in our truncation and dress
them in accordance with gauge symmetry, see Apps.~\ref{app:mSTIvert}
and \ref{trunc:qgl} for further details. Similarly we find that the
chiral symmetry breaking operator 
\begin{eqnarray}\label{eq:STI_operator_II}
 \bar q\, \left(\delta_{\mu\nu} +\left[\gamma_\mu,\gamma_\nu\right]\right)  D_\mu D_\nu \,q\,,  
\end{eqnarray}
contributes to $\mathcal{O}(\bar q A q)$ to the tensor
\begin{eqnarray}
 \tfrac{1}{2}{\cal T}^{(2)}_{\bar q A q}+{\cal T}^{(4)}_{\bar q A q}\ .
\end{eqnarray}
Since this is the most relevant operator in the phase of spontaneously
broken chiral symmetry we again include the corresponding contribution
to the two-quark-two-gluon vertex with gauge invariant
dressing. The explicit calculation of the dressing
  functions of the higher interactions to check the quantitative
  importance of deviations from the STI which are expected to occur
   below momenta of $\mathcal{O}(1$ GeV$)$ is deferred to future work. 

We want to stress at this point that if we would only take into
account a full basis for the quark-gluon vertex without the
corresponding gauge invariant partner tensor structures in the higher
vertices, we would see considerably different results. In
particular, the running coupling, see
Fig.~\ref{fig:running_couplings}, as defined from the dressing
function of the classical tensor structure, $z_{\bar q A q}^{(1)}$,
would deviate from the corresponding ghost-gluon running coupling at
considerably larger momenta. However, the degeneracy of the running
couplings defined from the different vertices at semi-perturbative
momentum scales is a consequence of gauge invariance. Hence we
conclude that the higher quark-gluon vertices are important for the
consistency of the truncation. Moreover, since diagrams that contain
the two-quark-two-gluon vertex have a different number of quark lines
than the ones that contain only classical vertices, we expect a
qualitative effect at finite chemical potential due to these higher
quark-gluon interactions. 

Note that for assessing the importance of the
different tensor structures one has to take into account not only
their relative strength but also their respective symmetry properties.
For example, 
simply extracting the relative strength from 
Fig.~\ref{fig:quarkgluonvertex_components_msp}, we would conclude that 
the operator in \eq{eq:STI_operator_I} seems to be the most 
important one by far. We find, however, that also the operator
in \eq{eq:STI_operator_II} is very important for the value of the
quark propagator mass function. This is explained by the fact that
\eq{eq:STI_operator_II} breaks chiral symmetry and contributes
therefore directly to the quark mass function.

\subsection{Quark propagator}

Next we discuss our solution for the quark propagator, see
Fig.~\ref{fig:quarkpropagator_msp}, for an earlier study
see e.g. \cite{Fischer:2003rp}, and the effect of different 
quark-gluon interactions. We find very convincing agreement with
results obtained in lattice QCD in the quenched approximation
\cite{Bowman:2005vx}, that are shown with dimensionful quantities
rescaled by a factor of $0.91$ to match the scale of \cite{Sternbeck:2006cg}
and \cite{Fischer:2008uz,FP}. 
However, some care is necessary when comparing our propagator to the lattice
results, since the quenched approximation in lattice simulations sets
the fermion determinant to unity, whereas we just used a quenched
gluon propagator. 

Apart from the classical tensor structure, the most important
contribution to the quark propagator stems from the tensor structures
$\tfrac{1}{2}{\cal T}^{(5)}_{\bar q A q}+{\cal T}^{(7)}_{\bar q A q}$
for $Z_q(p)$, and $\tfrac{1}{2}{\cal T}^{(2)}_{\bar q A q}+{\cal
  T}^{(4)}_{\bar q A q}$ for $M_q(p)$, where we find it necessary to
include the full momentum dependence of the corresponding dressing
functions $z_{\bar q A q}^{(i)}(p,q)$. 
It is only the combination of all these terms, including
their gauge invariant partner structures in the quark-gluon vertex equation
together with their momentum dependencies, that leads to the very good agreement with the
lattice propagator. In particular this concerns the wave function
renormalisation $Z_q(p)$ for small momenta, where an important
contribution stems from mesonic fluctuations in the infrared.
These fluctuations have been included
with functional methods, e.g. in
\cite{Jungnickel:1995fp,Berges:2000ew,Schaefer:2006ds,Braun:2009gm,Mitter:2013fxa,Alkofer:1993gu,Fischer:2007ze,Fischer:2008sp,Cloet:2008re}.
Restricting the discussion only to the relative importance of
quark-gluon vertex tensor structures, recent findings in
Dyson-Schwinger studies
\cite{Hopfer:2013np,Williams:2014iea,Aguilar:2014lha} agree with our, 
see also \cite{Alkofer:2008tt} for an earlier study. 
Moreover, our present findings suggest the inclusion of the
STI-consistent  higher quark-gluon interactions in future DSE-studies. 

Finally we want to point out that one crucial contribution to the
quark mass function comes from the addition to the flow of the Yukawa
coupling, $\partial_t\Delta h$, due to dynamical hadronisation, see \eq{eq:Delta1}. As soon as
one runs into the spontaneously broken phase of QCD, $\langle
\sigma\rangle\neq 0$, this term contributes to the quark mass function
as well via the relation $\partial_t\Delta M_q(0)\varpropto \langle
\sigma\rangle \partial_t\Delta h_\pi $. Momentum dependencies are very
important in $\partial_t\Delta M_q(p)$, since we expect this term to be
approximately zero for momenta larger than the chiral symmetry
breaking scale, see App.~\ref{sec:dynhad} for details. Similarly, we
had to include momentum dependencies in the remaining four-fermi
interactions that appear in the tadpole diagram. In the language of
dynamical hadronisation, chiral symmetry breaking in terms of the
quark mass function is then triggered by the additional term,
$\partial_t\Delta M_q$, see \eq{eq:Delta2}. This, however, is just due to the chosen
parameterisation of the four-fermi interaction in terms of mesons.
Without dynamical hadronisation, chiral symmetry breaking would be
driven by the tadpole diagrams containing the resonant
(momentum-dependent) four-fermi channel, which in turn is driven by
quark-gluon interactions.

\subsection{Gluonic vertices and running couplings}
\label{sec:runningcouplings}

\begin{figure*}
  \subfloat[Renormalisation group scale dependence of dimensionless
  four-fermi interactions, see App.~\ref{trunc:4f}, and bosonised
  $\sigma$-$\pi$ channel. Grey: respects chiral symmetry, blue: breaks
  $U(1)_A$, red: breaks $SU(2)_A$, magenta: breaks
  $U(2)_A$.\hfill\textcolor{white}{.}]{\includegraphics[width=0.48\textwidth]{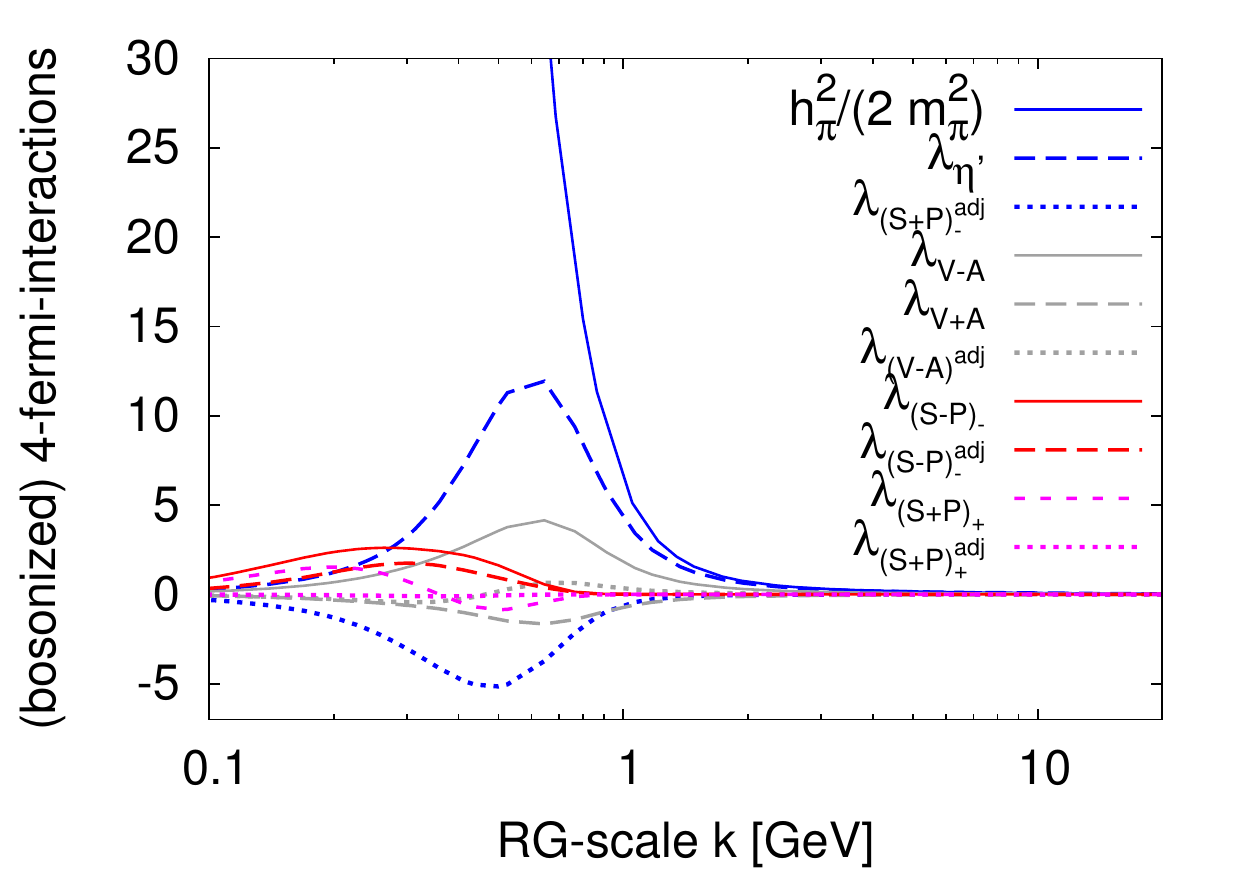}
      \label{fig:fourfermis_msp}} \hfill \subfloat[Independence of
    infrared model parameters from initial value and scale
    demonstrated for Yukawa coupling in a qualitative approximation,
    see also
    \cite{Braun:2014ata}.\hfill\textcolor{white}{.}]{\includegraphics[width=0.48\textwidth]{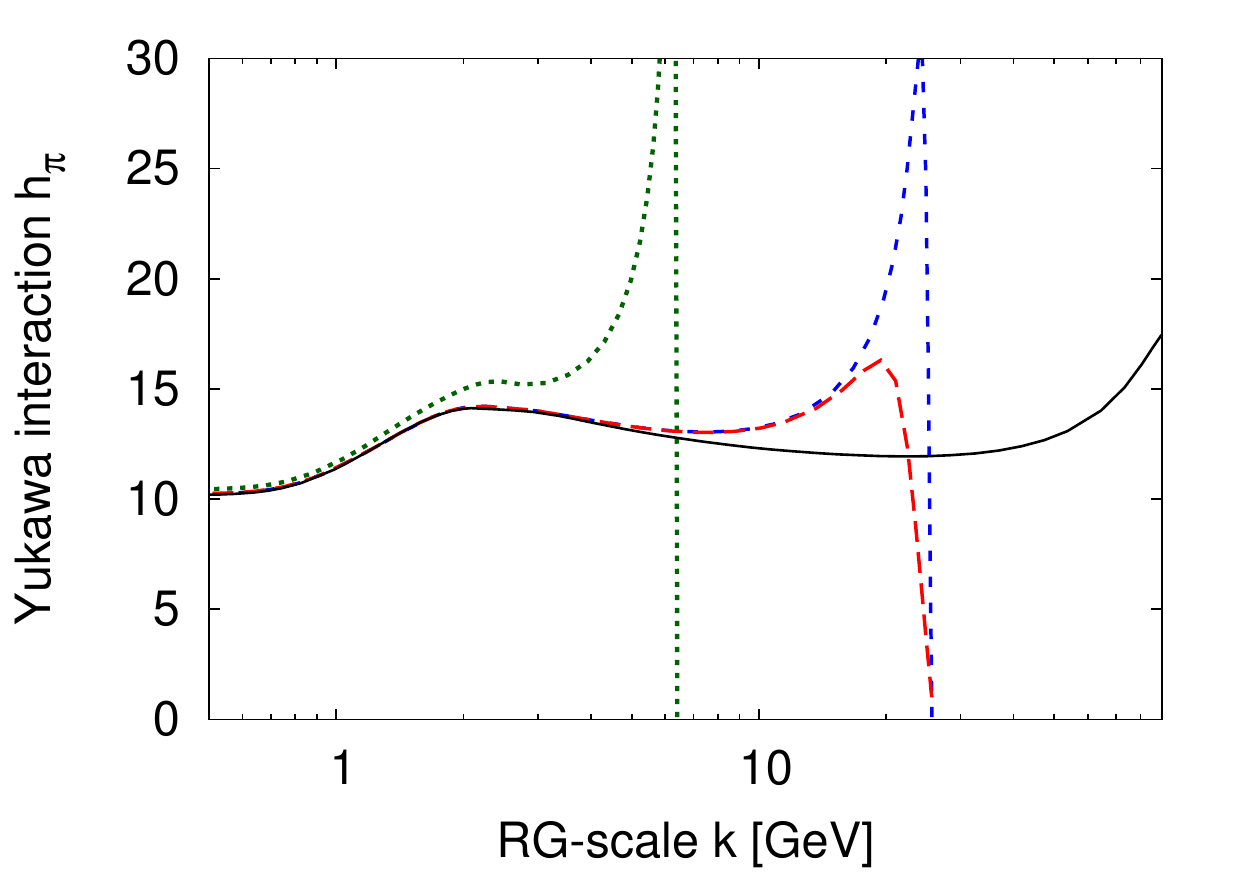}
    \label{fig:yukawaconvergence}}
  
  \caption{RG-scale dependence of four-fermi interactions and Yukawa coupling.}
\end{figure*}

From our calculated momentum-dependent QCD vertices, namely from the
quark-gluon, the ghost-gluon and the three-gluon vertex we can extract
running couplings. Following the detailed discussion in
App.~\ref{app:mSTI}, these running couplings are required to be
degenerate at (semi-)~perturbative momenta by means of Slavnov-Taylor
identities, whereas they will start to deviate in the non-perturbative
regime below momenta $\mathcal{O}(1$ GeV$)$.  Additionally, there is
no unique definition of a running coupling extracted from a particular
vertex in the non-perturbative regime. Here we define effective
running couplings that explicitly take the decoupling due to the
gluonic mass gap into account, illustrated exemplarily for the running
coupling extracted from the quark-gluon vertex evaluated at the
symmetric point,
\begin{align}
\label{eq:alphasAqq}
\alpha_{\bar q Aq}(p^2) &= \frac{\left(z_{\bar q A
      q}^{(1)}(p,q)\right)^2 }{4\pi }\Bigg\vert_{p^2=q^2=(p+q)^2}\,.
\end{align}
The running couplings from different vertices are shown in
Fig.~\ref{fig:running_couplings}.  Irrespective of the definition, all
running couplings coincide down to momenta of 4~GeV. This
underlines the fact that STI violations are negligible, without the
necessity to tune initial conditions as an implicit solution of the
STI as described in App.~\ref{app:mSTIinitial}. Even more, the
degeneracy of all running couplings at large momenta represents a
highly nontrivial statement about the consistency of our truncation,
in the sense that all important contributions in the semi-perturbative
regime have been consistently taken into account. Note that the very
good agreement of the quark-gluon and the ghost-gluon running coupling
is in part a consequence of the full momentum dependence which is
self-consistently taken into account in the quark gluon vertex. This
suggests that a similar improvement in the glue sector might lead to
an even better agreement of the three-gluon coupling with the two
other couplings. At low momenta, the gap in the gluon propagator
becomes important and we find a clear difference between the strength
of the various vertices. In particular, the 
three-gluon vertex drops considerably earlier, which is mainly 
due to the larger number of gluon legs.

\subsection{Relative importance of four-fermi channels}
\label{sec:4fermi}

As mentioned in the discussion of our truncation in
Sec.~\ref{sec:truncation}, we include a Fierz-complete basis with all
ten basis elements consistent with the $SU(N_c)_c\otimes
SU(N_f)_V\otimes U(1)_B$ symmetry to assess the importance of
different four-fermi interaction channels, see App.~\ref{trunc:4f} for
details on the choice of the basis. All four-fermi couplings are shown
in Fig.~\ref{fig:fourfermis_msp}, where in particular $h_{\pi}^2/2
m_{\pi}^2$ corresponds to $\lambda_{(S-P)_+}+\lambda_{(S+P)_-}$ and
$\lambda_{\eta'}$ to $\lambda_{(S-P)_+}-\lambda_{(S+P)_-}$.  We find
that the dynamics of spontaneous chiral symmetry breaking is almost
exclusively driven by the chirally symmetric four-fermi channel
$\lambda_{(S-P)_+}$, which corresponds to the quantum numbers of the
$\sigma,\pi,\eta$ and $a$-mesons.  However, this channel is split by
the presence of the 't~Hooft determinant coupling,
$\lambda_{(S+P)_-}$, such that only the $\sigma$-meson and pions
become very light. The dynamically created quark mass is already
sufficient to strongly suppress the $\eta'$-channel in comparison to
the resonant pion channel. Additional contributions due to the
$U(1)_A$-anomaly would lead to an even stronger suppression, see
\cite{Pawlowski:1996ch,Mitter:2013fxa}. Additionally,
  for sufficiently large initial scales, these anomalous
  contributions are suppressed relative to the contribution that
  orginates from the explicit symmetry breaking due to non-vanishing
  current quark masses. First checks indeed indicate that the anomalous
  contributions at a sufficiently initial large cutoff scale do not
  play a quantitatively important r\^ole, a more detailed study will
  be presented elsewhere. On the other hand, anomalous contributions 
  corresponding to fluctuations below the cutoff scale are already 
  taken into account by integrating the FRG running. This has also been
  demonstrated e.g. for the quantum mechanical anharmonic 
    oscillator \cite{Zappala:2001nv}. 
Therefore, all but the resonant pion
four-fermi channel constitute subleading contributions with a
quantitative effect of less than $5$ \%, see
Fig.~\ref{fig:4fermirebos}.  Independent of their relative strength,
the suppression of any of the four-fermi interactions is overcome by
the strength of $\alpha_s$ only in the non-perturbative regime of QCD
at scales of $\mathcal{O}(1$ GeV$)$.

In the light of these results it is sufficient to take into account only the
$(\sigma-\pi)$-channel, provided one uses a projection
obtained from a full basis to avoid ambiguities in the projection
procedure. Furthermore, note that in the purely fermionic theory all
ten channels diverge at the chiral symmetry breaking scale signaling
resonant quark-anti-quark states, as illustrated in
Fig.~\ref{fig:4fermirebos} for the $(\sigma-\pi)-$channel. These
divergencies are a consequence of ignoring momentum dependencies and
can be removed by dynamically hadronising only the
$(\sigma-\pi)$-channel. Nevertheless it would be interesting to also
bosonise other four-fermi channels to investigate the properties of the
corresponding bound states. Alternatively an investigation of the
momentum dependencies of the four-fermi interactions themselves is also
conceivable. As a word of caution, our statements about the relative
strength of four-fermi channels are only valid in the vacuum as in
particular finite chemical potential is expected to shift the relative
strength of four-fermi interaction channels.

Finally we want to mention that Fig.~\ref{fig:fourfermis_msp} captures
only the zero external momentum limit of the four-fermi interaction
channels. Although it was necessary to calculate the momentum
dependence of the $s$-channel momentum configuration for the
evaluation of the quark-propagator for this work, we postpone a thorough
discussion of the momentum dependence of the four-fermi interactions to
future publications. Here we only note that the effect of such
momentum dependencies on the remainder of the matter system is very
weak, since all but the $\sigma$-$\pi$ channel are very
weak. In the latter, on the other hand, we have implicitly taken
momentum dependencies into account via dynamical hadronisation.

Furthermore we want to stress that the dynamical hadronisation procedure
of introducing effective mesonic degrees of freedom introduces no model parameters
in the theory. Therefore, the infrared physics in terms of quark and mesons
are independent from the ultraviolet starting point and initial values. This
is demonstrated explicitly for the Yukawa interaction between quarks and 
mesons in Fig.~\ref{fig:yukawaconvergence}, where different initial scales
and initial values for the Yukawa interaction have been chosen, but where all
trajectories converge towards the same trajectory in the IR.

\subsection{Mechanism of chiral symmetry breaking}
\label{sec:chiralsymmbr}

As outlined in the introduction a proper understanding of the mechanisms
of confinement and chiral symmetry breaking is a crucial step towards
a quantitatively reliable approach to the phase diagram of QCD at
finite chemical potential. Here we comment on the mechanism of
chiral symmetry breaking from the point of view of the matter system.

In
\cite{Gies:2001nw,Gies:2002hq,Gies:2003dp,Gies:2005as,Braun:2005uj,Braun:2006jd,Braun:2011pp}
a simple picture for chiral symmetry breaking in quenched QCD was put
forward. In their analysis the IR fixed points in the four-fermi
interactions are destabilised if the gauge coupling exceeds a critical
coupling $\alpha_\text{crit}$ and as a result the four-fermi coupling
becomes singular. Although the argument is qualitatively
correct, in quenched QCD the picture is not so simple, as the drop of
the gauge coupling at small momenta, see Fig.~\ref{fig:alpha_msp}, lets the quark sector become
subcritical again. This was discussed as one possible scenario in
\cite{Braun:2006jd}, but is confirmed here as the actual physical
situation. In Fig.~\ref{fig:alpha_msp}, we show the different running
couplings and the critical gauge coupling. Since the gauge coupling
decreases below the critical coupling for decreasing momenta, it is merely
the area above the critical value line which is decisive for the occurrence
of chiral symmetry breaking.

Our findings indicate that an approach where the vertex strength of all
tensor structures of the quark-gluon vertex is subsummed in an
enhanced strength of the classical tensor structure lacks quantitative
precision. Using
such an enhanced quark-gluon vertex in our calculation would lead to much
too large contributions in the four-fermi sector, from gluonic box
diagrams which grow like $\alpha^2_{\bar q A q}$. Taking into
account different tensor structures approximately corresponds to a sum
of contributions $\simeq \sum_i \alpha_i^2$, if we denote the running
couplings associated to different components of the quark-gluon vertex
as $\alpha_i$ and neglect cross terms, whereas the enhanced vertex
from the single channel approximation contributes as the square of the sums
$\simeq (\sum_i\alpha_i )^2$ in the four-fermi box diagram.

Finally, we briefly discuss the mechanism of chiral symmetry breaking
which is at work in our framework. Here, chiral symmetry breaking is
driven by four-fermi interactions. In a framework of dynamical
hadronisation this is reflected in the corresponding contributions to the
Yukawa coupling/quark mass. Therefore, our calculation requires
significantly less vertex strength in the quark-gluon vertex in order
to see chiral symmetry breaking compared to the required strength in
single channel approximation as described above. In our framework, including
just the classical tensor structure in the quark-gluon vertex leads to
qualitatively albeit not quantitatively correct results. This is
mainly due to the contributions from the tensor structure $\tfrac{1}{2}{\cal
  T}^{(5)}_{\bar q A q}+{\cal T}^{(7)}_{\bar q A q}$ in the
quark-gluon vertex and its gauge invariant completion, see the
discussion in Sec.~\ref{sec:quark} and App.~\ref{trunc:qprqglmom}.

\section{Summary and conclusions}

In the present work we have investigated spontaneous chiral symmetry
breaking in quenched continuum QCD. The only relevant couplings are
those of QCD: the strong coupling $\alpha_s$ and the current quark
masses which are fixed in the perturbative regime. In particular this
allows us to compute the quark propagator in excellent agreement with
corresponding results from lattice QCD.

The functional renormalisation group analysis presented here 
uses a vertex expansion that goes qualitatively beyond the
approximation level used so far in continuum methods. On the one hand
advanced approximations have been used in sub-systems such as the pure 
glue sector and the low-energy matter sector. 
On the other hand we have, for the first time,
introduced a complete basis of four-fermi interactions in the
$s$-channel as well as the full quark-gluon vertex with all its
momentum-dependencies and tensor structures. The latter has been
linked to higher order quark-gluon interactions via modified
Slavnov-Taylor identities. These higher order terms are also important
for the convergence of the results, which emphasises the
necessity of an expansion scheme based on gauge-invariant operators.  
The quantitative reliability has been discussed in a detailed analysis of
the systematic errors.

The transition from the quark-gluon to the hadronic phase is smoothly
done by means of dynamical hadronisation. This allows to monitor the
emergence of composite mesonic operators as dynamical degrees of
freedom at low energies. We have also investigated the relative
importance of different four-fermi interaction channels. Here we find
that a single channel approximation with $\sigma$ and $\vec \pi$ is
sufficient to induce spontaneous chiral symmetry breaking on a
semi-quantitative level. This fact together with the small width of
the strongly-correlated transition region from the quark-gluon regime
to the hadronic regime, see also \cite{Braun:2014ata}, can be used to 
systematically improve the reliability of low-energy effective models, see
\cite{Pawlowski:2010ht,Haas:2013qwp,Herbst:2013ufa,Pawlowski:2014aha}.

The present computation is currently being extended to full dynamical
QCD, for first investigations see \cite{Braun:2014ata}, and to finite
temperature and density. Our analysis of the matter sector should also
give access to the large density regime, provided the higher
fermionic interactions including fluctuating baryons are monitored
accordingly.

\acknowledgments We thank R.~Alkofer, J.~Braun, C.S.~Fischer,
L.~Fister, T.~K.~Herbst, M.~Hopfer, M.Q.~Huber, F.~Rennecke, B.-J.~Schaefer,
L.~von~Smekal, R.~Williams and A.~Windisch for discussions. This work
is supported by the Helmholtz Alliance HA216/EMMI, the grant
ERC-AdG-290623, the FWF through Erwin-Schr\"odinger-Stipendium
No. J3507 and the BMBF grant 05P12VHCTG.

\appendix

\section{Modified Slavnov-Taylor identities}
\label{app:mSTI}
In the presence of the regulator terms the standard Slavnov-Taylor
identities (STI) are modified (mSTI).  Here we briefly discuss these
modifications and their implications following \cite{Pawlowski:2005xe},
a more detailed study will be presented elsewhere. A very concise form of
these identities is found in a formulation with the auxiliary
Nakanishi-Laudrup field $\lambda$ with
\begin{align}\label{eq:NL}
    e^{-\tfrac{1}{2} \int_x (\partial_\mu A^a_\mu)^2}\to 
\int\mathcal{D}\lambda\, e^{ -\tfrac{1}{2}
  \int_x \partial_\mu A^a_\mu\, \lambda^a- \tfrac{\xi}{2} \int_x \partial_\mu
  \lambda^a\lambda^a }\,.
\end{align}
where the full classical action $S=S_{\rm QCD}+S_{\rm gf}+S_{\rm ghost}$
is invariant under the BRST transformations
\begin{align}\nonumber 
  {\mathfrak s} \Phi & = (D_\mu c\,,\, -\tfrac{1}{2} f^{abc} c^b\,c^c\,,\,\lambda^a\,,\, -c\,
  q\,,\, \bar q\, c\,, 0,0,\dots)\,,\\[2ex] 
{\mathfrak s}
  \lambda & =0\,.
\label{eq:BRST}
\end{align}
with $\Phi$ as defined in \eq{eq:Phi}. In \eq{eq:BRST} we have assumed that
all the composite fields introduced in $\Phi$ are colourless. The
introduction of the auxiliary field $\lambda$ leads to ${\mathfrak
  s}^2 \phi=0$. The cutoff terms are not invariant under the BRST
transformations in \eq{eq:BRST} and the standard STI is modified. It
reads in a compact way
\cite{Ellwanger:1994iz,D'Attanasio:1996jd,Igarashi:2001mf,Pawlowski:2005xe}
\begin{align} \label{eq:mSTI}
  \int \0{\delta\Gamma_k}{\delta
    Q_{\phi_i}}\0{\delta\Gamma_k}{\delta\phi_i}= \int \, R_{k,\phi_n
    \phi_i}\, \0{\delta^2\Gamma_k }{\delta Q_{\phi_i}\delta \phi_j}\,
  G_{\phi_j\phi_n}\,,
\end{align}
where we have added a BRST source term 
\begin{align}\label{eq:BRST-soucre}
\int Q_{\phi_i} ({\mathfrak s} \phi)_i\,,
\end{align} 
to the path integral, see \cite{Pawlowski:2005xe} for more details and
further references. The sums in \eq{eq:mSTI} run over all species of
fields including internal indices. 

\subsection{Initial conditions for vertices}
\label{app:mSTIinitial}
In the limit $k\rightarrow 0$ the left hand side of \eq{eq:mSTI} vanishes and
we arrive at the standard STI: the derivative of $\Gamma$ with respect
to $Q_\phi$ generates the (quantum) BRST transformations that act
linearly on the fields via the derivative of $\Gamma$ with respect to
$\phi$. For perturbative momenta $p$ this gives the standard relations between the
renormalisation factors of the vertex functions at $k=0$, that is 
\begin{equation}\label{eq:pertBRST}
 z^{2}_{\bar c Ac}(p)=z^{2}_{\bar q Aq}(p)=z^{2}_{A^3}(p) = z_{A^4}(p)\,,
\end{equation}
corresponding to degenerate running couplings in the perturbative
regime. \Eq{eq:pertBRST} entails that in QCD the parameters of the
theory are given by the power-counting relevant mass parameters of the
quarks (dimension one), one (marginal) coupling, $\alpha_s$, and the
unobservable (marginal) wave function renormalisations of the fundamental fields.

In turn, for $k\neq 0$ the simple relations in \eq{eq:pertBRST} are
in general lost and the loop term on the right hand side of \eq{eq:mSTI} leads to
modifications. In terms of power-counting the most relevant modification is the
occurrence of a longitudinal gluon mass parameter $m_{k,L}^2$ in the gluon
propagator that vanishes for $k\to 0$. Perturbatively it relates to a
transversal mass parameter $m_{k,\bot, \rm pert }^2=m_{k,L,\rm pert}^2$.
Non-perturbatively this relation does not hold anymore, as we have 
a non-vanishing transversal mass gap in Landau gauge, for more details 
see \cite{FP}. In the present work we do not solve the mSTIs for the vertices
explicitly and also avoid the necessity of discussing the decoupling
of transversal and longitudinal parameters. We take the very good realisation
of \eq{eq:pertBRST} in our results, i.e. Fig.~\ref{fig:running_couplings}, as an
indication that the effects of the right hand side of \eq{eq:mSTI} at
the initial scale $\Lambda$ are either small or become unimportant during
the evolution of flow equation.

\subsection{STI-invariant vertices} 
\label{app:mSTIvert}
In the perturbative regime, the relations between the gluonic vertices
can be obtained with the help of the renormalisation group invariant
covariant derivative 
\begin{align}\label{eq:rencov}
D_{\mu} = \partial_\mu - i\, \sqrt{4 \pi \alpha_{s} \bar Z_{A}} A_\mu \,,
\end{align} 
with the renormalised gauge field $A_\mu$. Using the corresponding
field strength tensor
\begin{align}\label{eq:Fren}
  F_{\mu\nu}= \0{i}{  \sqrt{ 4 \pi \alpha_{s}} } [ D_{\mu}\,,\,D_{\nu}]\,,
\end{align}
in the effective action leads to gluonic vertices that are consistent
with gauge invariance for momenta $p\approx k$ down to $\mathcal{O}($1 GeV$)$. 
By using $\alpha_{s,k}(p)$ and $Z_{A,k}(p)$ these relations could be
made true for a larger range of momenta, which, however, is not necessary
due to the locality of the flow. In any case, such relation cannot be found
for the non-perturbative regime, since the perturbative mSTIs are in general
only valid for the longitudinal part of the vertices which deviate from their
transversal parts at non-perturbative momenta as it is for example the case for the
gluon mass gap. Consequently the vertices have to be computed separately 
and do not follow from the mSTI at low momenta. However, the gluonic vertices 
and ghost-gluon vertex gain a potentially significant non-trivial 
momentum-dependence and non-classical tensor structures only in the deep infrared 
where their contributions decouple due to the mass gap in the gluon propagator. 
In the present work we use this as a justification for approximating the
four-gluon vertex by the three-gluon vertex.

On the other hand, the quark-gluon vertex potentially gets a
significant non-trivial momentum-dependence and non-classical tensor
structures (see \eq{eq:tensorqAq}) below momenta of $\mathcal{O}(1$
GeV$)$, where chiral symmetry breaking is triggered. To take these
effects into account we study the STI-consistent version of 
the most important tensor structures
\begin{align}\label{eq:z57}
z_{\bar q A q}^{(5)}  (\slashed p+\slashed q) (p-q)^\mu\,, \qquad z_{\bar q A q}^{(7)} 
\tfrac{1}{2}[\slashed p,\slashed q]\gamma^\mu\,,  
\end{align}
see
\Fig{fig:quarkgluonvertex_components_msp}.  It can be shown that terms
proportional to $z_{\bar q A q}^{(5)}$ and $z_{\bar q A q}^{(7)}$ are
derived from
\begin{align} \label{eq:mSTI-57}
  z_{\bar q A q}^{(\rm av)}\, \bar q\, \gamma_5 \gamma_\mu \,\epsilon_{\mu\nu\rho\sigma} D_\nu
  D_\rho D_\sigma \,q\,,
\end{align}
with 
\begin{align}
z_{\bar q A q}^{(7)} = z^{(\rm av)}_{\bar q A q} \,,\qquad z_{\bar q A q}^{(5)} = \tfrac{1}{2} z^{(\rm av)}_{\bar q A q}\,, 
\end{align} 
valid for constant $z_{\bar q A q}^{(\rm av)}$. In \eq{eq:mSTI-57} we have used that
the mSTI-consistent extension of the $z^{(5)}_{\bar q A q}, z^{(7)}_{\bar q A q}$-momentum structure in \eq{eq:z57} is
\begin{align}\label{eq:mSTI-pre57}
\tfrac{1}{4}   \bar q\, D_\mu\,\{ [ \gamma_\mu\,,\,\gamma_\nu]\,,\,  \dr\} \,  D_\nu\, q\,,
\end{align} 
The tensor structure \eq{eq:mSTI-pre57} only projects on the $z^{(5)}_{\bar q A q}$ and
$z^{(7)}_{\bar q A q}$-terms.  After some algebra \eq{eq:mSTI-pre57} can be rewritten as \eq{eq:mSTI-57} 
by using 
\begin{align} 
 \tfrac{1}{4} z_{\bar q A q}^{(\rm av)}\,  \{[\gamma_\mu\,,\,\gamma_\nu]\,,\, \gamma_\rho\} =
  \epsilon_{\mu\nu\rho\alpha}\gamma_5 \gamma_\alpha\,.
\end{align}
With \eq{eq:Fren} we get
\begin{align}\label{eq:mSTI57final}
z^{(\rm av)}_{\bar q A q} \tfrac{i}{4} \sqrt{ 4 \pi \alpha_{s} } \, \bar q\, \gamma_5
  \gamma_\mu \,\epsilon_{\mu\nu\rho\sigma} \{ F_{\nu\rho}\,,\,
  D_\sigma\} \,q\,.  
\end{align}
Hence, as long as 
\begin{align} 
z^{(7)}_{\bar q A q}-2 z^{(5)}_{\bar q A q}=0\,,
\end{align} 
the quark-gluon vertex tensor structures \eq{eq:z57} follow from the
single term \eq{eq:mSTI57final}, see Fig.~\ref{fig:STI}.  This term is derived from the mSTI
which is valid down to the semi-perturbative regime. We conclude that within a
self-consistent approximation one also has to take into account the
related quark-gluon scattering vertices ($\bar q A^2 q$ and $\bar q
A^3 q$) derived from \eq{eq:mSTI57final}. A more detailed study
will be published separately. Finally, it is noteworthy that the tensor structure \eq{eq:mSTI57final}
also plays a crucial r\^ole in the so-called transverse Ward-Takahashi identity for the quark-gluon
vertex, see \cite{Kondo:1996xn} and e.g. \cite{Qin:2013mta,Qin:2014vya,Aguilar:2014lha} for applications.

\section{Truncation}
\label{app:truncation}

In this section we discuss in detail the different constituents of the
truncation scheme introduced in Sec.~\ref{sec:truncation}.
\subsection{Yang-Mills sector}

\subsubsection{Propagators}
\label{trunc:gpr} 

The only external input which is required in our calculation are the
pure gauge propagators. In Landau gauge, the inverse gluon propagator can be
parameterised as
\begin{align}\label{eq:glueprop_dressing}
  \Gamma_{A^2}^{\mu\nu}(p) &= Z_A(p)p^2 \Pi^{\mu\nu}_T(p)\,,
\end{align}
where $\Pi^{\mu\nu}_T(p)$ denotes the transverse projector 
\begin{align}
 \Pi^{\mu\nu}_T(p)&=\left(\delta^{\mu\nu}-\frac{p^\mu p^\nu}{p^2}\right) \, .
\end{align}
In addition to the gluon propagator we also encounter the ghost
propagator 
\begin{eqnarray}
 \Gamma_{\bar c c}(p) &= Z_c(p) p^2 \,.
\end{eqnarray}
in the equations for the pure gauge vertices. 

We stress that the matter sector computation does not rely on a
particular propagator input, but can use any available propagators.
This includes RG-scale and momentum-dependent propagators as provided
by FRG calculations \cite{Fischer:2008uz,FP} or just
momentum-dependent input such as lattice Yang Mills propagators in
minimal Landau gauge from \cite{Bowman:2004jm,Sternbeck:2006cg} where
the former will be used here, see Fig.~\ref{fig:gluonprop}. Taking an
external input for the propagators automatically sets the scale of the
theory and, apart from the bare quark mass, no parameters remain in
the perturbative regime, where we set our initial condition.

\subsubsection{Vertices}
\label{trunc:3g}

We approximate the ghost-gluon vertex, the three-gluon vertex and the
four-gluon vertex with their classical tensor structures and a 
momentum-dependent dressing, $z_X$ for $X\in\{\bar c A c,A^3,A^4\}$,
\begin{align}
  &\Gamma_{\bar c A c}(p_1,p_2)_{\mu}^{abc} = z_{\bar c A
    c}(\bar p) Z_c(\bar p) Z_A^{1/2}(\bar p) \left[\imag gf^{abc} q_\mu\right]\
  ,\nonumber\\[2ex]
  & \Gamma_{A^3}(p_1,p_2)_{\mu\nu\rho}^{abc} = \nonumber\\[1ex]
  & \qquad z_{A^3}(\bar p) Z_A^{3/2}(\bar p) \bigg[\imag f^{abc}
  \Big\{(p_2-p_1)_\rho
  \delta_{\mu\nu}+ \text{ perm.}\Big\}\bigg]\ ,\nonumber\\[2ex]
  & \Gamma_{A^4}(p_1,p_2,p_3)_{\mu\nu\rho\sigma}^{abcd} =\nonumber\\[1ex]
  & \qquad z_{A^4}(\bar p) Z_A^{2}(\bar p)
  \left[f^{iab}f^{icd}\delta_{\mu\rho}\delta_{\nu\sigma}
    + \text{ perm.}\right]\ .
\end{align}
Here the $p_i$, denote the momenta and we approximate the dressing
functions $z_X$ as functions of one average momentum $\bar p \equiv
\sqrt{\sum_i p_i^2/\sum_i}$.  To project onto the dressing functions
we multiply each gluon leg with the corresponding transversal
projector. Therefore the projection on the ghost-gluon vertex dressing
is uniquely defined whereas we contract the three-gluon vertex equation
additionally with $\delta_{\mu\nu}p_{2,\rho}-\delta_{\nu\rho}p_{2,\mu}$. The
four-gluon vertex is approximated from the three-gluon vertex via
\begin{equation}
\label{eq:4gluonapprox}
 z_{A^4}(\bar p) = z_{A^3}^2(\bar p),
\end{equation}
which leads to an approximate agreement of the three-gluon running coupling with the ghost-gluon and
quark-gluon running coupling down to
$\mathcal{O}($1 GeV$)$ and is still expected to improve with an improved
momentum resolution of the glue sector, see the discussion in Sec.~\ref{sec:runningcouplings}.

\subsection{Matter Sector}\label{trunc:qprqglmom}
 
\subsubsection{Quark propagator}
\label{trunc:qpr}
We parameterise the inverse dressed quark propagator with two dressing
functions as
\begin{align}\label{eq:quarkprop_dressing}
 \Gamma_{\bar qq}(p) &= Z_q(p) \left(\imag\slashed{p} + M_q(p)\right)\ ,
\end{align}
where 
\begin{align}\label{eq:gamma+clifford}
  \{\gamma_\mu\,,\,\gamma_\nu\} =2\delta_{\mu\nu} \id \,,\qquad
  \gamma_\mu^\dagger =\gamma_\mu\,,\qquad \gamma_5=
  \gamma_1\gamma_2\gamma_3\gamma_4\,.
\end{align}
Setting the current quark mass, $M_q(20\text{ GeV})= 1.3$ MeV, 
is related to the value of the pion mass, see App.~\ref{trunc:qm}.
Apart from the purely mesonic sector of our truncation, the full
momentum dependence of the quark propagator is fed back into the
equations for all other vertices. In the quark-meson sector, as
described in Sec.~\ref{trunc:qm}, such an approximation leads to an
overestimation of the suppression of loops containing quarks via the
quark mass function. The resulting effect will most likely be an
underestimation of the order parameter $\langle\sigma\rangle$ since
the quarks drive the order parameter to larger values in the
quark-meson model.

\subsubsection{Quark-gluon interactions}
\label{trunc:qgl}
In Landau gauge, a basis for the quark-gluon vertex is given by the
eight tensor structures
\begin{equation}
\begin{split}
[{\cal T}^{(1)}_{\bar q A q}]^\mu(p,q)&=\gamma^\mu\ ,\\[2ex]
[{\cal T}^{(2)}_{\bar q A q}]^\mu(p,q)&=-\imag(p-q)^\mu\ ,\\[2ex]
[{\cal T}^{(3)}_{\bar q A q}]^\mu(p,q)&=-\imag(\slashed p-\slashed q)\gamma^\mu\ ,\\[2ex]
[{\cal T}^{(4)}_{\bar q A q}]^\mu(p,q)&=\imag(\slashed p+\slashed q)\gamma^\mu\ ,\\[2ex]
[{\cal T}^{(5)}_{\bar q A q}]^\mu(p,q)&=(\slashed p+\slashed q) (p-q)^\mu\ ,\\[2ex]
[{\cal T}^{(6)}_{\bar q A q}]^\mu(p,q)&=-(\slashed p-\slashed q) (p-q)^\mu\ ,\\[2ex]
[{\cal T}^{(7)}_{\bar q A q}]^\mu(p,q)&=\tfrac{1}{2}[\slashed p,\slashed q]\gamma^\mu\ ,\\[2ex]
[{\cal T}^{(8)}_{\bar q A q}]^\mu(p,q)&=-\tfrac{\imag}{2}[\slashed p,\slashed q](p-p)^\mu\ ,
\end{split}\label{eq:tensorqAq}
\end{equation}
where $p$ ($q$) denotes the momentum of the (anti-)quark. It is
important to note that the tensor structures ${\cal T}^{(2)}_{\bar q A
  q}$, ${\cal T}^{(3)}_{\bar q A q}$, ${\cal T}^{(4)}_{\bar q A q}$ and
${\cal T}^{(8)}_{\bar q A q}$ break chiral symmetry and are only
created in the spontaneously broken phase. Our final Ansatz for the quark-gluon vertex
is then
\begin{align}\label{eq:quarkgluon-basis}
  \Gamma_{\bar q A q}(p,q) &= -\imag Z_{q}(\bar p)Z_A^{1/2}(\bar p) \nonumber\\[1ex]
  & \times \sum\limits_{i} \frac{z^{(i)}_{\bar q A q}(p,q)}{\bar p
    ^{\, n_i}}[{\cal T}^{(i)}_{\bar q A q}]^\mu(p,q)\ ,
\end{align}
where $\bar p ^{\, n_i}$ is the average momentum and $n_i$ is chosen
such that $z^{(i)}_{\bar q A q}(p,q)$ is dimensionless.

As remarked in Sec.~\ref{sec:quark} a sensible truncation scheme
should also include a set of higher order operators to complete it consistent
with the STI.  We find that the most important contributions to
non-classical tensor structures in the quark-gluon vertex stem from
terms of the form
\begin{eqnarray}
 \bar q\, T_{\mu\nu}D_\mu D_\nu q\ ,\qquad
 \bar q\, T_{\mu\nu\rho}D_\mu D_\rho D_\nu q\ ,
\end{eqnarray}
where the first (second) contribution breaks (respects) chiral
symmetry. In momentum space these yield contributions to the action of
the form
\begin{align}
 \mathcal{O}(\bar q Aq): &\ \bar q(p)\Big\{T_{\mu\nu}(\imag q_\nu)+T_{\nu\mu}(-\imag p_\nu)\nonumber\\
				    &+T_{\mu\nu\rho}(-q_\nu q_\rho) +T_{\nu\mu\rho}(- p_\nu p_\rho) + T_{\rho\nu\mu}(p_\rho q_\nu)\Big\}\nonumber\\[1ex] 
							&\  \times \left[-\imag g A_\mu(-p-q)\right] q(q) \ ,\nonumber\\[2ex]
 \mathcal{O}(\bar q A^2q):&\ \bar q(p)\Big\{T_{\mu\nu}\nonumber\\
				    &+T_{\rho\mu\nu}(-\imag p_\rho) +T_{\nu\mu\rho}(\imag(r+q)_\rho) + T_{\nu\rho\mu}(\imag q_\rho)\Big\}\nonumber\\[1ex] 
							&\  \times \left[-\imag g A_\nu(-p-q-r)\right]\left[-\imag g A_\mu(r)\right]q(q) \ ,\nonumber\\[2ex]
 \mathcal{O}(\bar q A^3q):&\ \bar q(p)T_{\mu\nu\rho}\left[-\imag g A_\mu(-p-q-r+s)\right]\nonumber\\[1ex]
							&\  \times \left[-\imag g A_\rho(s)\right]\left[-\imag g A_\nu(r)\right]q(q) \,,
\end{align}
where $g$ is to be understood as $g=\sqrt{4\pi\alpha(\bar p) Z_{A}(\bar p)}$ everywhere. In particular, 
we find that the dominant contributions to order $D^2$ and $D^3$ correspond to the tensor structures
\begin{align}
  T_{\mu\nu} & = \delta_{\mu\nu} +\left[\gamma_\mu,\gamma_\nu\right]\ ,\nonumber\\[2ex] 
  T_{\mu\nu\rho} & = \{[\gamma_\mu,\gamma_\nu],\gamma_\rho\}\ .
\end{align}
In terms of quark-gluon tensor structures from \eq{eq:tensorqAq} these are proportional to the linear combinations
\begin{align}
    &\tfrac{1}{2}{\cal T}^{(2)}_{\bar q A q}+{\cal T}^{(4)}_{\bar q A q}\ ,\nonumber\\[2ex]
    &\tfrac{1}{2}{\cal T}^{(5)}_{\bar q A q}+{\cal T}^{(7)}_{\bar q A q}\ ,   
\end{align}
see \ Fig.~\ref{fig:quarkgluonvertex_components_msp}.

Therefore we set for consistency reasons $z_{\bar q A q}^{(5)}=\frac{1}{2} z_{\bar q A
  q}^{(7)}$ and $z_{\bar q A q}^{(2)}=z_{\bar q A  q}^{(4)}$ in all equations.
 The dressing of the corresponding higher order operators
is then simply identified with that of the corresponding quark-gluon
vertex dressings $z_{\bar q A q}^{(7)}$ and $z_{\bar q A q}^{(4)}$
respectively where we additionally use the RG-invariant ansatz as in all other vertices, 
e.g.
\begin{align}
  &\bar q(p)Z_{q}(\bar p) z_{\bar q A q}^{(4)}(\bar p) T_{\mu\nu}\left[-\imag \sqrt{4\pi\alpha(\bar p) Z_{A}(\bar p)} A_\mu(r)\right] \nonumber\\[1ex]
		   & \ \times \left[-\imag \sqrt{4\pi\alpha(\bar p) Z_{A}(\bar p)} A_\nu(-p-q-r)\right]q(q)  \ ,
\end{align}
with $\bar p = \sqrt{(p^2+q^2+r^2+(p+q+r)^2)/4}$.

\subsubsection{Four-fermi interactions}
\label{trunc:4f}
Here we discuss a basis for the four-fermi interactions where $(S\pm P)$/
$(V\pm A)$ denotes the scalar-pseudoscalar/vector-axialvector Dirac
structure, the subscript denotes the flavor structure and the
superscript the color structure. Omitted sub-/superscripts are
to be understood as singlet contributions.

A basis for the $U(2)_L\times U(2)_R$ symmetric four-fermi interactions
is given by \cite{Jaeckel:2002rm}, see also \cite{Braun:2011pp} for a
review,
\begin{align}\label{eq:fourfermi_sym}
  \mathcal{L}_{(\bar q q)^2}^{(S-P)_+^{\phantom{adj}}}&=(\bar q T^0
  q)^2\!-
  \!(\bar q \gamma^5 T^f q)^2\!-\!(\bar q \gamma^5 T^0  q)^2\!+\!(\bar q T^f q)^2\nonumber\\[2ex]
  \mathcal{L}_{(\bar q
    q)^2}^{(V-A)_{\phantom{-}}^{\phantom{adj}}}&=(\bar q \gamma^\mu
  T^0 q)^2\!+\!(\bar q \gamma^\mu\gamma^5 T^0 q)^2\nonumber\\[2ex]
  \mathcal{L}_{(\bar q
    q)^2}^{(V+A)_{\phantom{-}}^{\phantom{adj}}}&=(\bar q
  \gamma^\mu T^0 q)^2\!-\!(\bar q \gamma^\mu\gamma^5 T^0 q)^2\nonumber\\[2ex]
  \mathcal{L}_{(\bar q q)^2}^{(V-A)_{\phantom{-}}^{\text{adj}}}&=(\bar
  q \gamma^\mu T^0 T^a q)^2\!+\!(\bar q \gamma^\mu\gamma^5 T^0 T^a
  q)^2\,.
\end{align}
We denote the generators of flavor $U(1)$ and $SU(2)$ by $T^0$ and
$T^{f}$ whereas $T^a$ are the generators of color $SU(3)_c$. Note,
that the obvious choice $(S-P)_+^\text{adj}$ instead of
$(V-A)^\text{adj}$ is not linearly independent of $(S-P)_+$ and
$(V+A)$ and therefore $(V-A)_{\phantom{-}}^{\text{adj}}$ has to be
considered.

There are two four-fermi interactions which break the axial $U(1)_A$ but
are symmetric under $U(1)_V\times SU(2)_L\times SU(2)_R$
\begin{align}\label{eq:fourfermi_anom}
  \mathcal{L}_{(\bar q q)^2}^{(S+P)_-^{\phantom{adj}}}=&(\bar q T^0
  q)^2\!-
  \!(\bar q \gamma^5 T^f q)^2\!+\!(\bar q \gamma^5 T^0  q)^2\!-\!(\bar q T^f q)^2\nonumber\\[2ex]
  \mathcal{L}_{(\bar q q)^2}^{(S+P)_-^{\text{adj}}}=&(\bar q T^0 T^a q)^2\!-\!(\bar q \gamma^5 T^f T^a q)^2\!\nonumber\\[1ex]
  &\quad+\!(\bar q \gamma^5 T^0 T^a q)^2\!-\!(\bar q T^f T^a q)^2\, ,
\end{align}
where the first corresponds to the 't~Hooft determinant
\cite{'tHooft:1976fv}. For applications it is
convenient to introduce the linear combinations
\begin{align}
  \mathcal{L}_{(\bar q q)^2}^{(\pi)}&=\mathcal{L}_{(\bar q
    q)^2}^{(S-P)_+} +
  \mathcal{L}_{(\bar q q)^2}^{(S+P)_-}=2 (\bar q T^0 q)^2-2(\bar q \gamma^5 T^f q)^2\nonumber\\[2ex]
  \mathcal{L}_{(\bar q q)^2}^{(\eta')}&=\mathcal{L}_{(\bar q
    q)^2}^{(S-P)_+} -
  \mathcal{L}_{(\bar q q)^2}^{(S+P)_-}=2 (\bar q T^f q)^2-2(\bar q \gamma^5 T^0 q)^2
  \end{align}
with quantum numbers corresponding to $(\sigma-\pi)-$ and $(\eta- a)-$
meson exchange channels.

Since the $SU(2)_L\times SU(2)_R$ symmetry is only approximate and
explicitly broken to $SU(2)_{L+R}$ we additionally take into account
the tensor structures
\begin{align}\label{eq:fourfermi_su2}
  \mathcal{L}_{(\bar q q)^2}^{(S+P)_+^{\phantom{adj}}}=&(\bar q T^0
  q)^2\!+\!(\bar q
  \gamma^5 T^f q)^2\!+\!(\bar q \gamma^5 T^0  q)^2\!+\!(\bar q T^f q)^2\nonumber\\[2ex]
  \mathcal{L}_{(\bar q q)^2}^{(S+P)_+^\text{adj}}=&(\bar q T^0 T^a
  q)^2\!+\!(\bar q
  \gamma^5 T^f T^a q)^2\nonumber\\[1ex]
  &\quad\!+\!(\bar q \gamma^5 T^0 T^a q)^2\!+\!(\bar q T^f T^a q)^2\ ,
\end{align}
which break $SU(2)_{A}$. Finally there are two
basis elements which break $SU(2)_{A}$ as well as $U(1)_{A}$
\begin{align}\label{eq:fourfermi_su2_anom}
  \mathcal{L}_{(\bar q q)^2}^{(S-P)_-^{\phantom{adj}}}=&(\bar q T^0
  q)^2\!+\!(\bar q
  \gamma^5 T^f q)^2\!-\!(\bar q \gamma^5 T^0  q)^2\!-\!(\bar q T^f q)^2\nonumber\\[2ex]
  \mathcal{L}_{(\bar q q)^2}^{(S-P)_-^{\text{adj}}}=&(\bar q T^0 T^a
  q)^2\!+
  \!(\bar q \gamma^5 T^f T^a q)^2\nonumber\\[1ex]
  &\quad\!-\!(\bar q \gamma^5 T^0 T^a q)^2\!-\!(\bar q T^f T^a q)^2\ .
\end{align}
Consequently a basis that respects $U(1)_V\times SU(2)_V$ consists
of ten elements and the Ansatz for the full four-fermi vertex is given by
\begin{align}\label{eq:4fermi-basis}
  \Gamma_{(\bar q q)^2,k}(p_1,p_2,p_3) &= Z^2_{q,k}(0) \sum\limits_{i}
  \frac{\lambda^{\ }_{i,k}(s)}{k^2}\mathcal{L}_{(\bar q q)^2}^{i}\ ,
\end{align}
where the sum runs over these $10$ tensor structures. We investigated
the momentum dependencies in the four-fermi interactions for three
momentum configurations corresponding to pure $s$-,$t$- and
$u$-channel momentum configurations on the basis of the given solution
at zero external momentum. For example for the $s-$channel we consider
$p_1=p_2=-p_3=-p_4=p$ corresponding to $s=4 p^2$.

\subsubsection{Quark-meson system: LPA$'$ approximation}
\label{trunc:qm}

We parameterise the inverse meson propagators as
\begin{align}
  \Gamma_{\sigma^2/\vec\pi^2,k}(p) &= Z_{\pi,k} \left(p^2 +
    m_{\sigma/\pi,k}^2\right)\ .
\end{align}
In the chirally symmetric phase the approximation $Z_\sigma\approx
Z_\pi$ is exact, whereas the deviations in the broken phase are
suppressed by the comparably large mass of the sigma-meson $m_\sigma$.
The mass terms can be absorbed into the definition of the effective
mesonic potential and will be discussed there. Additionally we neglect
the momentum-dependence of $Z_\pi$ which has been shown to be a
quantitatively reliable approximation \cite{Helmboldt:2014iya}. As a
consequence only the anomalous dimension
\begin{align}
 \eta_\pi &= -\frac{\partial_tZ_\pi}{Z_\pi}\ ,
\end{align}
appears in any of the flow equations. 

We perform a Taylor expansion of the effective mesonic potential in
$\rho$ \cite{Pawlowski:2014zaa}
\begin{align}
 V(\bar\rho\equiv Z_\pi \rho) &= \sum\limits_{j=0}^6\frac{v_j}{j!}(\bar\rho-\bar\rho_0)^j\ ,
\end{align}
with $\bar\rho_{0}\equiv Z_\pi\rho_0$ and $\rho_0$ scale-independent
such that $\bar\rho_{0}$ becomes the minimum of the effective
potential at $k\rightarrow 0$.  At this order of the Taylor expansion
we see convergence and a comparison to a calculation on a discrete
grid in $\rho$ yields perfect agreement.  The meson masses are
obtained from this potential as
\begin{align}\label{eq:meson_masses}
 m_\pi^2  &= V'(\bar\rho_0)\ ,\nonumber\\[2ex]
 m_\sigma^2  &= V'(\bar\rho_0)+2\bar\rho_0V''(\bar\rho_0)\ .
\end{align}
Therefore the value of the pion mass depends directly
on the expansion point $\bar\rho_0$ which, in turn, is directly proportional
to the current quark mass $M_q(\Lambda)$. In our case we choose
$M_q(20\text{ GeV})= 1.3$ MeV such that $m_\pi$ takes the physical 
value of $135$ MeV.

We consider only one Yukawa interaction of the $\sigma$--$\pi$ tensor
structure $\mathcal{L}_{(\bar q q)^2}^{(\pi)}$. 
Hence, integrating out the mesonic fields leads to
contributions to the four-fermi interaction $\mathcal{L}_{(\bar q q)^2}^{(\pi)}$. 
In other words,
the total coupling of $\mathcal{L}_{(\bar q q)^2}^{(\pi)}$ is a sum of the 
explicit four-fermi interaction and the part stored in the quark-meson sector of the
theory. The distribution of these fluctuations is done with dynamical
hadronisation explained in App.~\ref{sec:dynhad}.

The chirally symmetric Yukawa interaction reads
\begin{align}\nonumber 
 &  \int_{p_1,p_2} z_{\bar q\phi q}(p_1,p_2) \bar Z_\pi^\frac{1}{2}(p_1+p_2)
\bar Z_q^\frac{1}{2}(p_1) \bar Z_q^\frac{1}{2}(p_2) \\[1ex]
& \hspace{3cm}\times \phi(p_1+p_2)\, \bar q(p_1) \tau
  q(p_2)\,,
\label{eq:YukInt}\end{align}
where $\tau=(T^0,\gamma_5 \vec T)$. Reducing this to the $s$-channel with $p_1=p_2=p$ leads to 
\begin{align}\nonumber 
 &  \int_{p} z_{\bar q\phi q}(p,p) \bar Z^\frac{1}{2}_\pi(2 p) 
\bar Z^\frac{1}{2}_q(p) \bar Z^\frac{1}{2}_q(p)  \\[1ex]
& \hspace{3cm}\times \phi(2 p)\, \bar q(p) \tau
  q(p)\,.
\label{eq:YukInt1}\end{align} 
In our calculations we use the renormalised Yukawa coupling $h_{\pi}$,
with $\bar Z_{\pi}(2p) \equiv  Z_{\pi,k}(0)$,  $\bar Z_{q}(p) \equiv \bar Z_{q,k}(0)$
and
\begin{align}\label{eq:Yukcoup}
h_\pi(2 p) = 2 z_{\bar q\phi q}(p,p)\,, 
\end{align}
where $2p$ is the momentum of the mesonic field $\phi(2p)$.
Furthermore we ignore momentum dependencies as well as field dependencies
in $h_\pi$.

This parameterisation of the quark-meson model is termed LPA$'$
approximation and has been shown to be capable of approximating the
full momentum dependence very well
\cite{Helmboldt:2014iya}. Furthermore, it has been found that the
effect of higher meson quark interactions that stem from a possible
field-dependence in the Yukawa interaction would yield a decrease of
the the order of $10$ \% in the chiral condensate
\cite{Pawlowski:2014zaa}.

\section{Stability of the truncation}
\label{app:results_stab}
\begin{figure}
  \includegraphics[width=0.48\textwidth]{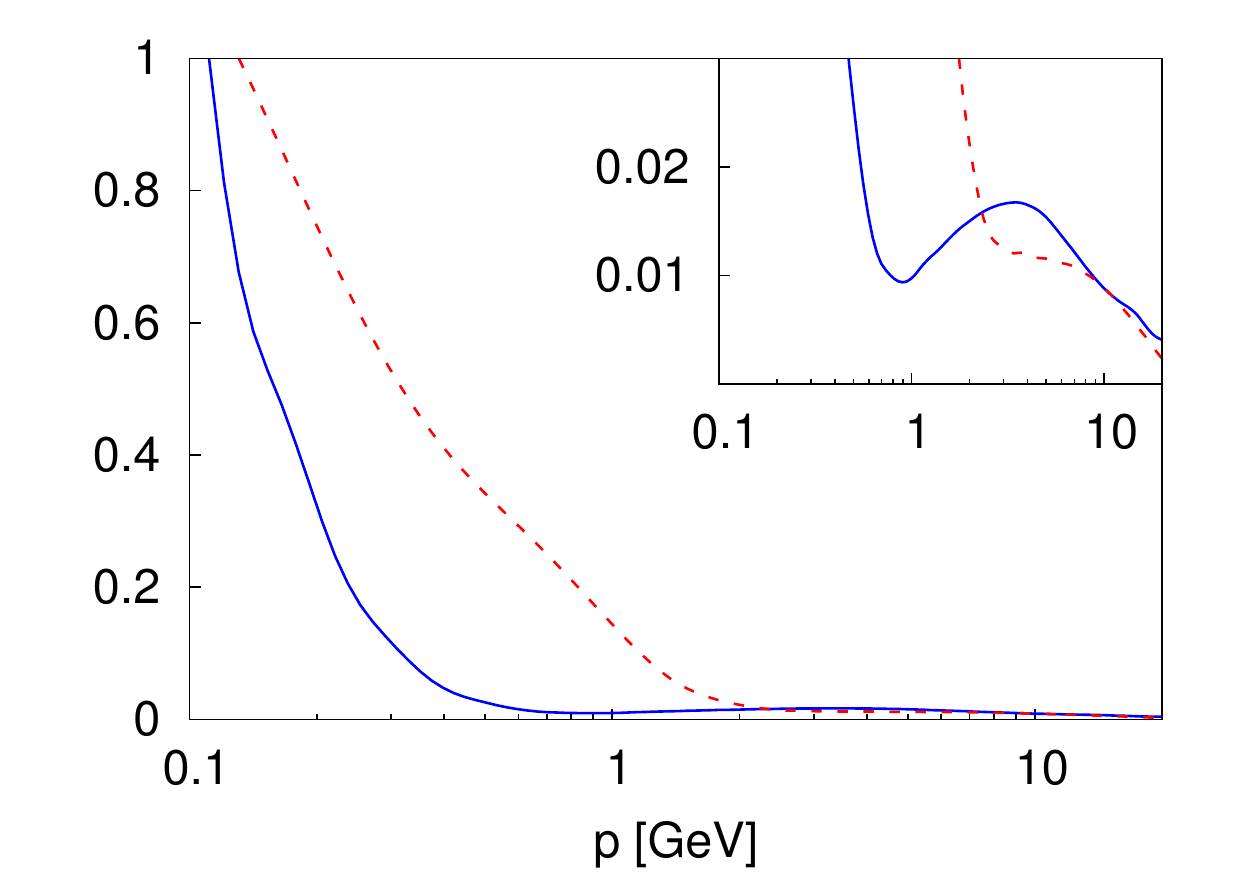}
  \caption{Normalised difference $(z^{(7)}_{\bar q A q}-2
    z^{(5)}_{\bar q A q})/z^{(7)}_{\bar q A q}$ (blue solid) and 
    $(z_{\bar c A c}-z^{(1)}_{\bar q A q})/z_{\bar c A c}$ (red dashed). 
    Small values indicate that the mSTI is applicable for
    constraining transversal tensor structures.\hfill\textcolor{white}{.}}
\label{fig:STI}
\end{figure}
Here we give a detailed analysis of the systematic errors and hence of
the stability of the current truncation. To this end we briefly
summarise the vertex structures taken into account: in the pure gauge
sector the classical tensor structures of all primitively divergent
correlation functions have been considered. The quark propagator and
the quark-gluon vertex - as the essential interface coupling between
glue and matter sector - have been included with full
momentum-dependencies and all tensor structures. Additionally, higher
quark-gluon interactions as obtained from a gauge invariant extension
of non-classical tensors structures in the quark-gluon vertex have
been taken into account. A Fierz-complete basis has been used for the
four-fermi couplings in an $s$-channel approximation. Moreover, meson
propagators and quark-meson Yukawa interactions as well as higher
order mesonic correlation functions in the scalar--pseudo-scalar
$s$-channel are included.\\[-1ex]

{\it Gluonic interactions:}  

As the momentum dependence of the Yang-Mills vertices has been found
to be rather small at the relevant momentum scales
\cite{Fister:2011uw,Fister:Diss,Huber:2012kd,Pelaez:2013cpa,%
  Blum:2014gna,Eichmann:2014xya,Binosi:2014kka,Gracey:2014mpa,Cyrol:2014kca},
we approximated the momentum dependence of these vertices only with
one variable at the symmetric point, which is expected to be a good
approximation.

For the three-gluon vertex, DSE studies show that the effect of
additional tensors structures is small \cite{Eichmann:2014xya}.
Our largest systematic
error concerns therefore the four-gluon vertex which has been determined via
STIs from the three-gluon vertex keeping only the classical tensor
structure. While this certainly works well in the semi-perturbative
regime, below $\mathcal{O}($1 GeV$)$ deviations are to be expected as well as
contributions from other tensor structures, see \cite{Cyrol:2014kca}.
Indeed, the three-gluon vertex
running coupling deviates already earlier from the other running
couplings, indicating some missing vertex strength, see
\Fig{fig:alpha_msp}.\\[-1ex]

{\it Quark-gluon interactions:} In the matter sector the quark-gluon
vertex was fully taken into account. However, based on an analysis of
the relative strength of the different tensor structures we only fed
back the ${\cal T}^{(4)}_{\bar q A q}$ and ${\cal T}^{(7)}_{\bar q A
  q}$ as the dominant chiral symmetry breaking and chiral tensor
structures. We have checked the quantitative convergence of this
approximation at the example of the flow equations of the quark
propagator and the quark-gluon vertex itself.

Higher quark-gluon interactions follow from the quark-gluon vertex
using the modified Slavnov-Taylor identities (mSTIs) discussed in
App.~\ref{app:mSTI}. 
The applicability of the mSTIs, which constrain only 
the longitudinal part of any correlation function, relies on identifying them with their
transversal counterparts. At non-perturbative momenta $\mathcal{O}($1 GeV$)$
the connection between transversal and longitudinal parts is lost, with
the running couplings as obtained from different vertices as a prominent example,
see comparison of $z_{\bar q A q }^{(1)} \varpropto \sqrt{\alpha_{\bar q A q}}$ with 
$z_{\bar c A c } \varpropto \sqrt{\alpha_{\bar c A c}}$ in Fig.~\ref{fig:STI}. 
In their regime of applicability the mSTIs provide therefore relations 
between different tensor structures, most prominently this leads to
$z^{(7)}_{\bar q A q}-2 z^{(5)}_{\bar q A q}\approx 0$, see
App.~\ref{app:mSTIvert}. In Fig.~\ref{fig:STI} we show the normalised
difference $(z^{(7)}_{\bar q A q}-2 z^{(5)}_{\bar q A
  q})/z^{(7)}_{\bar q A q}$ as obtained from the vertex equation with
only the classical tensor structure inserted on the right hand side.
In this case the mSTI is fulfilled even better than for the 
strong running coupling down to very low momenta  $\mathcal{O}($0.5 GeV$)$.
We take this as a justification for approximating the dressing and momentum dependence of the higher
quark-gluon interactions by the solution of the semi-perturbative mSTI
that relates them to the tensor structure $\tfrac{1}{2}{\cal
  T}^{(5)}_{\bar q A q}+{\cal T}^{(7)}_{\bar q A q}$. \\[-1ex]

{\it Quark interactions:} We have taken into account a complete Fierz
basis for the four-fermi interaction and used $s$-channel
approximations for all tensor structures. All higher purely fermionic
vertices in the $s$-channel of the scalar--pseudo-scalar interaction
are included. Their contribution beyond meson exchange
(eight-fermion interaction) is small, see \cite{Braun:2014ata}.  This
observation is additionally supported by the fast
convergence of expansions in powers of the mesonic field that is found
in the quark-meson model and in dynamical QCD,
at vanishing temperature \cite{Pawlowski:2014zaa,Helmboldt:2014iya}.  Concerning the
quark-meson interactions, we neglected higher contributions due to
field-derivatives of the Yukawa interaction which have been found to
be of the order of $10$ \% \cite{Pawlowski:2014zaa,Braun:2014ata}. 
A more detailed study will be presented elsewhere.

In the equation for the quark propagator, momentum dependencies of the
four-fermi interactions play a quantitative r\^ole via the tadpole
diagrams. We have implemented this momentum dependence
via one momentum variable using a symmetric projection.
Furthermore momentum dependencies in the meson sector have been
ignored which is justified by the success of the LPA$'$ approximation
\cite{Helmboldt:2014iya}. Additionally we have ignored the
backcoupling of the momentum dependence of the quark propagator in the
equation for the effective potential. We expect some effects due to
this approximation, which would mitigate the effect of
ignoring higher quark-meson interactions.

\section{Dynamical Hadronisation}
\label{sec:dynhad}

As already pointed out in the main text, the concept of dynamical hadronisation
\cite{Gies:2001nw,Pawlowski:2005xe,Floerchinger:2009uf} is
of crucial importance for the present application. In the form used
here it allows to exactly rewrite momentum channels of the four-fermi
interactions in terms of Yukawa couplings to an effective bosonic
exchange field. This corresponds to a Hubbard-Stratonovich
transformation in every RG-step, and is also called rebosonisation in
the present case of composite bosonic fields
\cite{Gies:2001nw}. Naturally, the bosonic field carries
the quantum numbers of the related four-fermi channel, and may be
interpreted as the corresponding meson or diquark field.

For simplicity we restrict ourselves in the following discussion to
the dynamical hadronisation of the sigma-pion-channel. Following
\cite{Pawlowski:2005xe} and in particular \cite{Braun:2014ata}, we start from
the path integral representation for $\Gamma_k[\Phi]$ in terms of the
fundamental superfield $\hat\varphi=(\hat A_\mu , \hat C,\hat{\bar C},
\hat q,\hat{\bar q})$
\begin{equation}
  e^{-\Gamma_k[\Phi]}=\int \mathcal{D} \hat{\varphi} \,
  e^{-S[\hat{\varphi}]-\Delta S_k[\hat{\phi}_k]+\frac{\delta (\Gamma_k
      +\Delta S_k)}{\delta \phi}
    (\hat{\Phi}_k-\Phi)+\Delta S_k[\Phi]}\,,
\end{equation}
with $\Delta S_k[\Phi]=\tfrac{1}{2}\Phi R_k \Phi$, where we introduced
a dynamical superfield $\hat \Phi_k=(\hat \varphi,\hat \sigma_k,
\hat{\vec{\pi}}_k)$ with expectation value $\Phi=\langle \hat
\Phi_k\rangle\equiv (\varphi,\sigma,\vec \pi)$. It is constructed
from the fundamental superfield $\varphi$ and scale-dependent
composite operators $\hat\phi_k=(\hat \sigma_k,\hat{\vec {\pi}}_k)$, whose flow
we define to be of the form
\begin{equation}
\label{eq:rebosfield}
\partial_t \hat\phi_k(r)=\partial_t A_k(r) \,(\bar q\tau q)(r)+
\partial_t B_k(r)\, \,\hat \phi_k(r)
\end{equation}
with $(\bar q\tau q)(r)= \int_l \bar
q(l) \tau q(r-l)$. 
The flow \eq{eq:rebosfield} is defined in momentum
space which will allow us to identify it with a specific momentum
channel in the four-fermi flow. Note also that the term multiplying
$\partial_t A_k$ involves only expectation values $q$ and $\bar
q$. The two coefficient functions $\partial_t A_k$ and $\partial_t
B_k$ appearing in \eq{eq:rebosfield} are so far undetermined and at
our disposal in the dynamical hadronisation. They specify the
RG-adaptive change of our field-basis. The scale-dependence of
$\hat\phi_k$ leads to additional contributions on the right hand side
of the flow equation compared to \eq{eq:floweq} which now takes the
form
\begin{align}\nonumber 
\partial_t|_\phi \Gamma_k[\Phi]=& \frac{1}{2}\,\text{Tr}\, 
\frac{1}{\Gamma^{(2)}_k+R_k}(\partial_t R_k+2 R_k \partial_t B_k)
\\[1ex]
&-\int_l \frac{\delta \Gamma_k}{\delta \phi} \Bigl[
\partial_t A_k(r)\, (\bar q \tau q)(r) +\partial_t B_k(r) \phi(r)\Bigr]\,.
\label{eq:DynHad}\end{align}
The second line on the right hand side account for the
scale-dependence of the composite fields. Together with the left hand
side they constitute a total derivative w.r.t.\ the logarithmic scale
$t$. In the present work we use $\partial_t A_k(r)$ to completely eliminate the
corresponding channel of the scalar--pseudo-scalar four-fermi interaction 
with $\lambda^{\ }_{\pi}\equiv \lambda^{\ }_{(S-P)_+}+
\lambda^{\ }_{(S+P)_-}$ and $\lambda^{\ }_{\pi}(s) = \lambda^{\
}_{\pi}(p,p,-p)$, see \eq{eq:4fermi-basis}, that is with $t=u=0$.
\begin{align}\label{eq:S-P0}
  \partial_t \lambda_{\pi}(s) ={\rm Flow}^{(4)}_{\pi}(s)
  - \partial_t A_k(2 p) h_\pi(2 p) \stackrel{!}{=} 0\,, 
\end{align} 
where ${\rm Flow}^{(4)}_{\pi}(s)$ stands for the
diagrams in the four-fermi flow. \Eq{eq:S-P0} leads to a vanishing flow of the
$s$-channel of the four-fermi coupling $\lambda_{\pi}$ and requires 
\begin{align}\label{eq:dtA}
  \partial_t A_k(2 p) = \0{{\rm
      Flow}^{(4)}_{\pi}(s)}{h(2 p)} \,,\qquad {\rm
    with}\qquad
s = 4 p^2\,,
\end{align}
which completely fixes $\partial_t A_k(2 p) $. Still, the second
rebosonisation function $\partial_t B_k(2 p)$ is at our disposal. It can
be used to improve the approximation at hand by distributing the
momentum-dependence of the rebosonised four-fermi channel between the
Yukawa coupling and the mesonic propagator. For example, when
considering the full momentum-dependence of the latter but only a
running, momentum-independent Yukawa coupling, the $\partial_t B_k$ can
be chosen such that this is an exact procedure. The discussion of the
general procedure is beyond the scope of the present work and will be
presented elsewhere. In the present case we resort to the simplest
option by using
\begin{align}\label{eq:dotB=0} 
\partial_t B_k\equiv 0\,.
\end{align}
As a consequence of \eq{eq:DynHad} together with \eq{eq:dtA} and
\eq{eq:dotB=0} we get additional contributions to the mesonic
anomalous dimension at vanishing momentum and the momentum-dependent
quark-meson coupling $h_\pi(2p)$,
\begin{align} 
  \partial_t  \Delta \eta_{\pi} &= 2\frac{V'(\bar\rho_0)}{h_{\pi}^2(0)}{\rm
    Flow}^{(4)}_{\pi}(0)\,,
  \nonumber\\[2ex]
  \partial_t \Delta h_{\pi} (2p) &= \frac{\Gamma_\pi^{(2)}(2 p)
  }{h_{\pi}(2p)}{\rm Flow}^{(4)}_{\pi}(s)\,.
 \label{eq:Delta1}\end{align}
The quark mass function is directly related to the quark meson
coupling, $M_q (p) = \langle\sigma\rangle h_\pi(2p)/2$. Moreover, in
the current approximation we use a constant $h_\pi=h_\pi(0)$ on the
right hand side of the flows. With
$\Gamma_\pi^{(2)}(0)=V'(\bar\rho_0)$ this leads us to
\begin{align}
  \partial_t \Delta h_{\pi} (0) &= \frac{V'(\bar\rho_0)}{h_{\pi}(0)}{
    \rm Flow}^{(4)}_{\pi}(0)\ ,\\[2ex]
  \partial_t \Delta M_q (p) &= M_q (p)
  \frac{V'(\bar\rho_0)}{h_{\pi}(0)^2}\frac{\bar \lambda_\pi(0)}{\bar
    \lambda_\pi(s)} \partial_t \bar\lambda_\pi(s)\, ,
  \label{eq:Delta2}\end{align} with \begin{align} \bar\lambda_\pi(s)=
  \int_{\Lambda_{\rm UV}}^k\0{d k'}{k'} {\rm Flow}^{(4)}_{\pi}(s)\,.
\end{align} In \eq{eq:Delta2} we have used that
\begin{align}\label{eq:4fh}
  \frac{\Gamma_\pi^{(2)}(2p)}{(h_{\pi}(2p))^2} \approx
  \frac{V'(\bar\rho_0)}{h_{\pi}(0)^2}\frac{\bar \lambda_\pi(0)}{\bar \lambda_\pi(s)}\, , 
\end{align}
up to higher order terms in the mesonic potential. In the present
approximation we have a better access to the momentum dependence of
$\bar \lambda_\pi$ than on that of $\Gamma_\pi^{(2)}$ and $h_\pi$.
Consequently, using \eq{eq:Delta2} minimises the error in our
computation of $M_q(p)$. For future work it would however be
preferable to calculate the momentum dependence of the right hand side
directly from momentum dependencies in the mesonic sector.

\begin{figure*}
	\centering
	\includegraphics[angle=90,height=0.45\textheight]{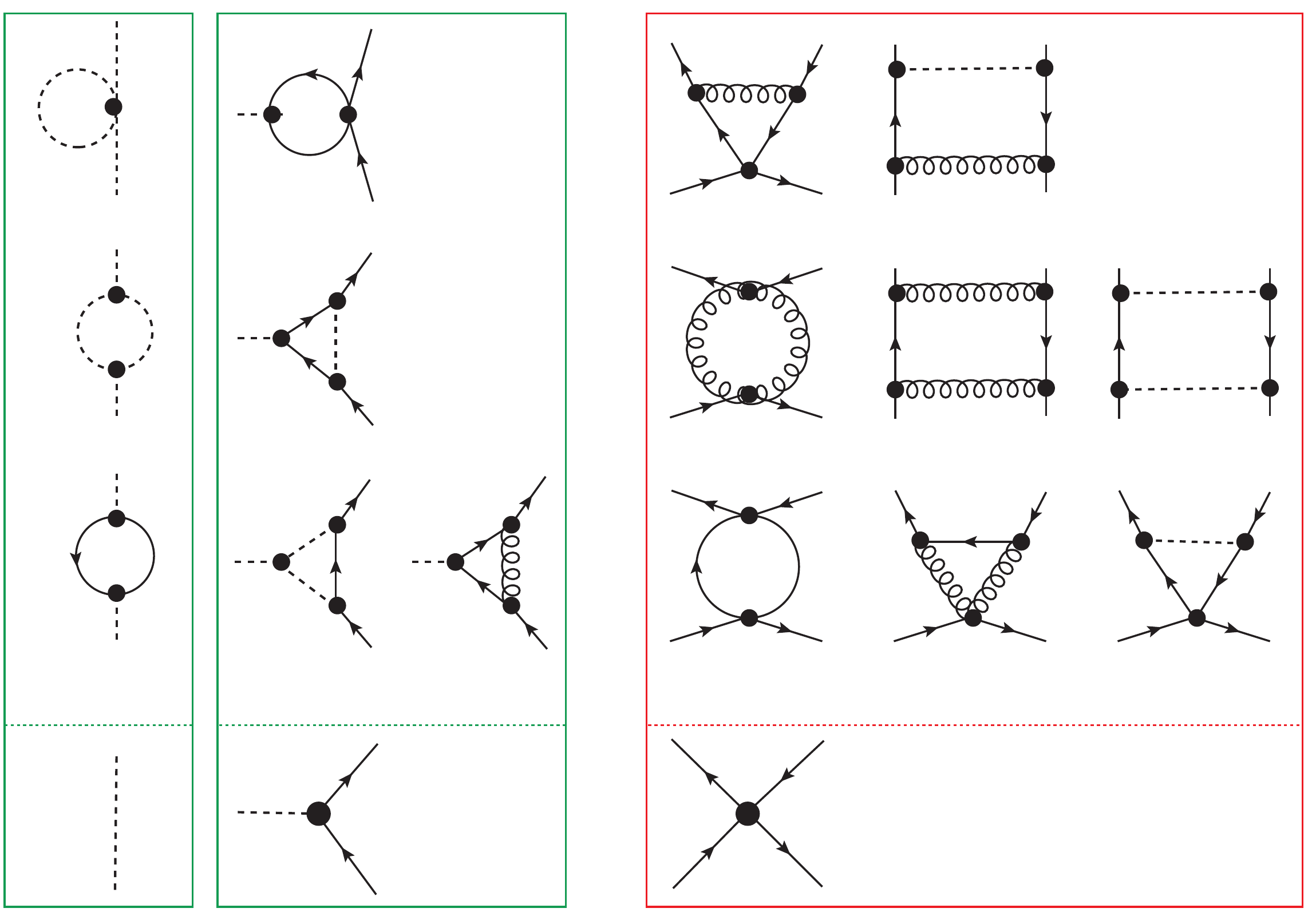}\vspace*{1.0cm}
	\includegraphics[angle=90,height=0.45\textheight]{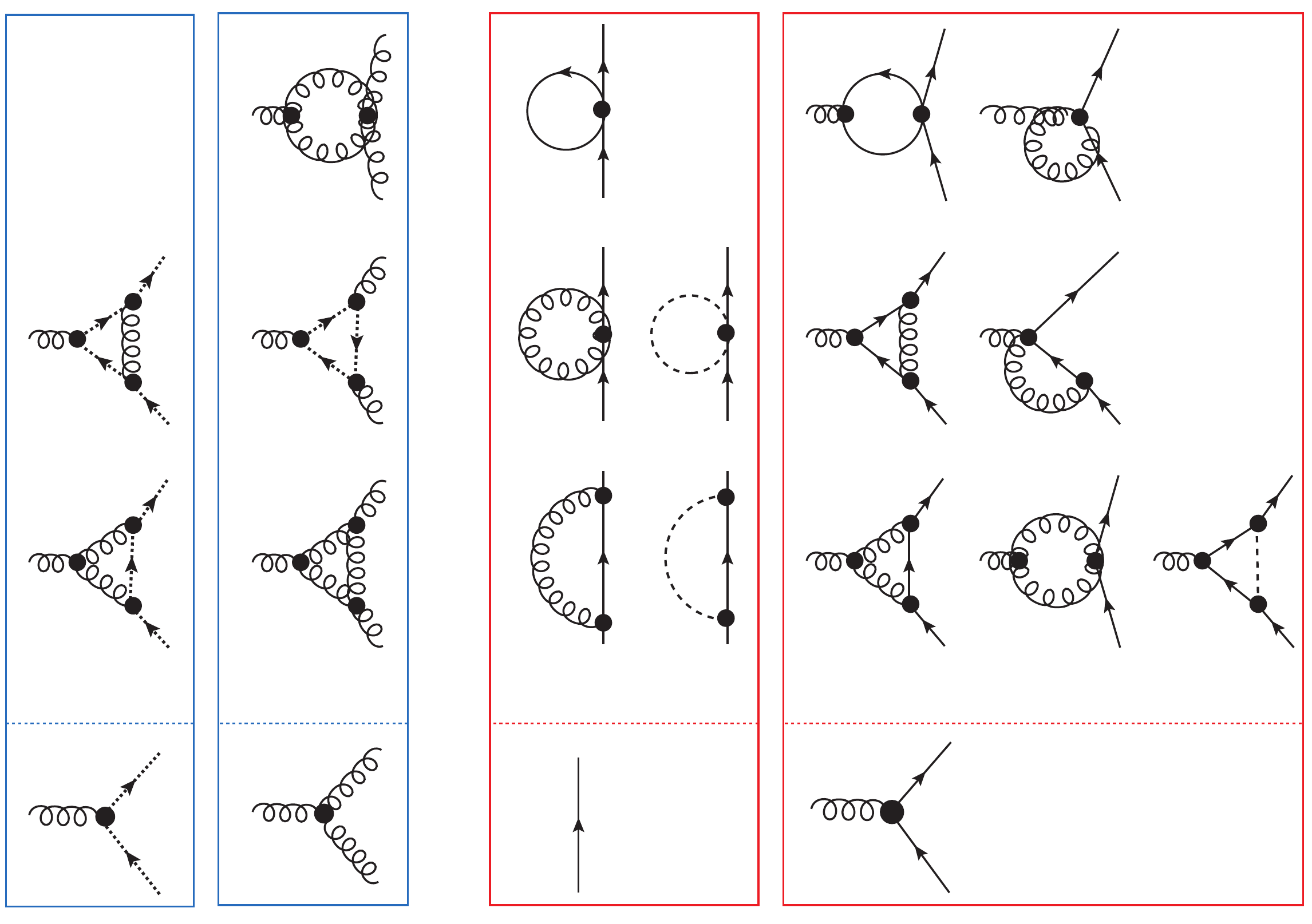}
	\caption{Types of diagrams that contribute to flow of propagators and vertices. Quark- (solid lines),
		gluon- (wiggly lines), ghost- (dotted lines) and meson- (dashed lines) propagators as well as
		the (1PI) vertices are dressed. Each diagram represents a sum of diagrams with (anti-)symmetric
		  permutations and regulator function inserted once in each internal propagator line. Symmetry
		  factors and signs are not shown for better readability. Not shown are the flows of ghost and
		    gluon propagators, see \cite{Fister:2011uw,Fister:Diss}, as well as of the effective potential.}
	\label{fig:flow_I}
\end{figure*}

\bibliography{../bib_master}

\end{document}